\documentclass[sigconf]{acmart}

\setcopyright{none} 
\acmYear{2018}

\usepackage{xcolor,colortbl}
\usepackage{hhline}
\usepackage{amsmath}
\usepackage{multirow}
\usepackage{amsfonts}
\usepackage{caption}
\usepackage{tikz}
\usepackage{booktabs}
\usepackage{pifont}
\usepackage{float}
\usepackage{enumitem}
\usepackage[T1]{fontenc}
\usepackage[titlenumbered,ruled,noline,linesnumbered]{algorithm2e}

\renewcommand{\vec}[1]{\mathbf{#1}}

\newcolumntype{a}{>{\columncolor[gray]{0.85}}l}
\newcolumntype{q}{>{\columncolor[gray]{0.85}}p}
\newcolumntype{b}{>{\columncolor{gray}{0.6}}l}

\begin{document}

\title{Effectiveness of Distillation Attack and Countermeasure on Neural Network Watermarking}

\author{Ziqi Yang\qquad Hung Dang\qquad Ee-Chien Chang}
\affiliation{%
  \institution{National University of Singapore}
}
\email{{yangziqi, hungdang, changec}@comp.nus.edu.sg}

\begin{abstract}
The rise of machine learning as a service and model sharing platforms has raised the need of traitor-tracing the models and proof of authorship.
Watermarking technique is the main component of existing methods for protecting copyright of models.
In this paper, we show that distillation, a widely used transformation technique, is a quite effective attack to remove watermark embedded by existing algorithms. 
The fragility is due to the fact that distillation does not retain the watermark embedded in the model that is redundant and independent to the main learning task. 
We design {\em ingrain} in response to the destructive distillation. It regularizes a neural network with an ingrainer model, which contains the watermark, and forces the model to also represent the knowledge of the ingrainer. Our extensive evaluations show that ingrain is more robust to distillation attack and its robustness against other widely used transformation techniques is comparable to existing methods.

\end{abstract}

\maketitle

\section{Introduction}

Machine learning (ML) models are becoming ubiquitous, powering an extremely wide variety of applications. As a consequence, they are being treated as conventional commodity softwares.
The model training could be outsourced, and the constructed models could be shared among various parties.
Cloud service providers, such as Google, Amazon, and Microsoft provide machine learning as a service which enables outsourcing machine learning and facilitates using third-party models.  Many companies, such as BigML, also operate exclusively on platforms that enable sharing and selling machine learning models.
As ML models are treated as an intellectual property, which can be sold or licensed, the needs of traitor-tracing and proof of authorship emerge. One can ``watermark'' the model during its training by implanting some secret into its parameters, which can then be used to prove one's authorship on the model\cite{embedding_watermarks, watermark_backdoor, adversarial_watermark, deepsigns, deepmarks}.  Compared to watermarking multimedia data, watermarking ML models poses new technical challenges since it is the model's functionality trained for a specific task that need to be protected.

A number of basic techniques are proposed in the literature for embedding secret information in neural networks, such as poisoning training data~\cite{backdoor_DL_posoning, ccs_ML_remembers, watermark_backdoor, adversarial_watermark, tamper_data}, modifying the training algorithm and retraining~\cite{badnets, trojanning_attack_NN, ccs_ML_remembers, adversarial_watermark, deepmarks, deepsigns}, or simply writing the secrets into the (e.g., least significant bits of) parameters after the training~\cite{ccs_ML_remembers}.
Some existing methods exploit the large capacity of neural networks in representing and memorizing random functions~\cite{memorizattion_DL, rethinking_generalisation_DL}, without damaging the model's accuracy~\cite{DL_robust_to_noise}.
Others exploit the massive unused capacity of the models' parameter space, which is largely redundant for the main classification function represented by the model~\cite{pruning_quantization}. 
Nonetheless, what is missing in the design of the existing watermarking techniques is that they are not designed strategically with countermeasures in mind. 
In fact, it is a common practice that models go through some transformations for memory, energy and computation optimization~\cite{compress_by_hashing, speeding_up_CNN_from_within, speedingNN_expansion, pruning_quantization, han2016eie, han_pruning,  distilling_DL, expand_nonlinear_CNN}, for fine-tuning with new data~\cite{fine_tune}, or for transfer learning~\cite{transfer_learning_cvpr, transfer_learning_survey}. 
These transformations can be actively used as attacks to remove watermarks in neural networks \cite{embedding_watermarks, watermark_backdoor, adversarial_watermark, deepsigns, deepmarks}.
Unfortunately, little has been studied about the impact of model transformations on the result of existing watermark embedding mechanisms. 

In this paper, we show that one of the widely used transformation techniques---distillation~\cite{distilling_DL}---is surprisingly a quite effective attack to remove the embedded watermarks. 
Distillation, as a type of compression techniques, uses the knowledge of the neural network to train a new model of smaller size.
We evaluate existing watermarking methods under distillation attack. 
The results empirically show that {\em all} the watermark information embedded in a neural network, using {\em any} of the existing methods, can be removed by distillation with negligible loss in the model's accuracy.
We perform a deep analysis on the results obtained. Our analysis reveals that existing methods which simply leverage the vast capacity of neural networks, leads to the embedded watermarks decoupled from the model's main functionality. More specifically, the sub-models or parameters which are responsible for memorizing the watermarks are almost independent from the part of the model which represents the main classification task.  As distillation is constrained to preserve the model's accuracy, the redundant information (which contains the watermarks, but does not contribute to the distillation's objective) will be lost.

In response to the fragility of existing watermarking methods against distillation, we design {\em ingrain}, a more robust watermarking method to counter distillation.
Ingrain essentially imbues the watermarks onto the predictions of the model on real (benign) data. 
The objective is to embed the watermark information into the same neural connections that are responsible for representing the main classification task.  
We achieve this by ingraining the watermarks in the main model's predictions through modifying the loss function of the classifier.  
We execute this in two steps.  First, we train an \textit{ingrainer} model exclusively on a watermark-carrier dataset (which is derived from the watermark information).  The ingrainer model has the same input-output format as the main classification model and contains all the watermark information (recoverable without any loss).  
In the second step, we use the loss function of the ingrainer as a regularization term for training the main classification model, so as to encourage the model to not only match the label for each training input, but also to match the output of the ingrainer on the same training data. 
Informally, this leads to a joint training of the classification model and the watermarking model, as opposed to independent training of the two objectives inside a model.  
By tuning the weight of the regularization term, we can trade-off the accuracy loss with the watermark robustness.

We extensively evaluated ingrain for various machine learning tasks and model architectures on multiple training datasets. The results show that, with acceptable accuracy loss, it improves the resistance of the embedded watermarks against distillation. 
For example, on one of the datasets, 30\% of the watermarks survive distillation with only 2\% loss in the classification accuracy.
Although ingrain is designed to counter distillation, it turns out that its robustness against other widely used transformations is comparable to existing methods.
This work highlights that even if the neural network is transformed into a different model via a strong attack (i.e., distillation), it is still possible to preserve watermark information if it is deeply ``ingrained'' in the model's main functionality.

\textbf{Contributions.} In summary, we make the following contributions in this paper.
\begin{itemize} [leftmargin=2em]

\item We empirically show that distillation is an effective attack which can remove all watermark information in neural networks embedded by existing methods. 

\item We perform a deep analysis of the robustness of existing methods.
We argue that minimizing the independence between the main task and the watermark embedding task inside a model can improve the robustness. Nonetheless, it intuitively harms the model's accuracy. We highlight the problem of studying the trade-off between embedding robustness and cost.

\item We design ingrain as an embedding technique to counter distillation. The watermark is embedded in such a way that it is correlated with the model's main task. We empirically show that ingrain can achieve better robustness against distillation. 
Besides, we also show that the robustness of ingrain against other widely used transformations (attacks) is comparable to existing methods.

\end{itemize}

\section{Machine Learning Model Lifecycle}\label{sec:background}

In this paper, we focus on supervised learning, more specifically, on
training classification models using (deep) neural
networks~\cite{DL_origin}.  The model is used to give predictions
to inputs provided by users.   

\subsection{Machine Learning Models}
\label{sec:ml}

A machine learning model, e.g., a neural network, encodes a general hypothesis function $F_w$ (with parameters $w$) which is learned from a training dataset with the goal of making predictions on unseen data.  The function maps some input space $\mathcal{X}$ to an output space $\mathcal{Y}$ (e.g., labels).  A neural network classification model $F_w(\vec{x})$ predicts the class for input $\vec{x} \in \mathcal{X}$ using a multi-layer network of basic non-linear activation functions (neurons) whose connections are weighted according to the model parameters $w$.  Each neuron obtains a number of activation signals from neurons in its preceding layer, whose importance weights are determined by their associated model parameters.  Then, it computes a non-linear activation function on the weighted sum of its input signals (plus a bias signal), and passes it to the neurons in the next layer.  An example of a widely used activation function is the rectified linear unit $\mathsf{relu}(z) = max(0,z),$ which we will also be using in our experiments.  In a classification model, a layer of normalized exponential function $\mathsf{softmax}(\vec{z})_i = \frac{\exp(z_i/T)}{\sum_j \exp(z_j/T)}$, where $T$ is the temperature, is added to the activation signals of the last layer to convert their arbitrary values into a vector of real values in $[0,1]$ that sum up to $1$.  Thus, the output could be interpreted as the probability that the input falls into each class.  

\subsection{Training \& Regularization} 
\label{sec:train} 

Let $\vec{x}$ be the data drawn from the underlying data distribution $p_x(\vec{x})$, and $y$ be the class of $\vec{x}$.
The training goal is to find the parameters $w$ such that the model $F_w$ is a good approximation of the mapping between 
every data point $(\vec{x},y)$ in the space $\mathcal{X}\times\mathcal{Y}$.
The accuracy of the model in this approximation is tested using a loss function $\mathcal{L} ( F_w(\vec{x}),y )$ that measures the difference between the class $y$ and the model's prediction $F_w(\vec{x})$.  
A common choice of $\mathcal{L}$ for classification models is the cross entropy loss function~\cite{murphy2012machine}.  
The training objective is to find a function $F_w$ which minimizes the expected loss.
\begin{equation}
L(F_w) = \mathbb{E}_{\vec{x}\sim p_x} [\mathcal{L}(F_w(\vec{x}), y)]
\end{equation}

It is intractable to accurately represent the actual probability function $p_x(\vec{x})$, but in practice, we can estimate it using samples drawn from it. These samples form the training set $D \subset \mathcal{X}$.
Hence, we can train the model to minimize the empirical loss over the training set $D$.
\begin{equation}
\label{classification_loss}
L_D(F_w) = \frac{1}{|D|} \sum_{\vec{x}\in D} \mathcal{L} (F_w(\vec{x}),y)
\end{equation}

Learning the optimal parameters is a non-linear optimization problem.  Algorithms used for solving this problem are variants of the gradient descent algorithm~\cite{gradient_descent}.  Stochastic gradient descent (SGD)~\cite{zhang2004solving} is a very efficient method that updates the parameters by gradually computing the average gradient on small randomly selected subsets of the training data.

The SGD algorithm navigates high dimensional space of the parameters to finds a local optimum of the loss function on the training data.  This could lead to an overfitted model which attains a very low prediction error on its training data, but fails to generalize well for unseen data.  Various {\em regularization} techniques have been introduced to mitigate this issue~\cite{DL_origin}.  These approaches try to prevent the parameter values from arbitrarily adapting to the training data.  This can be achieved by augmenting the training set, by adding a \textit{regularization term} $R(F_w)$ to the loss $L_D(F_w)$ for penalizing large parameters, or by randomly dropping the network connections during the training to prevent their complex co-adaptation during the training~\cite{dropout}.
The training process of a model can be summarized as to find a model $F_w$ that minimizes the following objective.
\begin{equation}
\label{regualarized_loss}
\mathcal{C}(F_w)=L_D(F_w)+\lambda R(F_w)
\end{equation}
where the regularization factor $\lambda$ controls the balance between the classification function and the regularization function.

\section{Watermark Embedding}
\label{sec:priorwork}

Given a watermark $S$, watermark embedding in a model $F_w$ refers to a process where the owner of $F_w$ embeds $S$ into $F_w$ in the training phase of the model, and later extracts $S$ from $F_w$ after its release to prove ownership on it.

In this section, we first formulate the watermark representation, and then introduce various embedding techniques in the literature. Next, we introduce distillation and other widely used transformation techniques of ML models which can be used as attacks to remove watermarks. 
We then analyze the robustness of existing embedding methods against distillation.

\subsection{Watermark Representation}

The watermark $S$ is typically represented as $n$ bits of information that carries the watermark in the model $F_w$. 
The owner can extract the $n$ bits of watermark information from $F_w$ as a proof of ownership.
Alternatively, the watermark can be encoded as $m$ predefined data-label pairs in a classification task with $k$ classes, where $m=\lceil \frac{n}{\lfloor\log_2(k)\rfloor} \rceil$. 
Let $D_S=\{\vec{x_s}_1, \vec{x_s}_2,\cdots,\vec{x_s}_m\}$ represent the predefined data sequence drawn from some data distribution $p_s$, and $Y_S=\{{y_s}_1,{y_s}_2,\cdots,{y_s}_m\}$ represent the predefined label sequence.
The embedding goal is to enforce a hidden function $F_w(\vec{x_s}_i)={y_s}_i$ in the model $F_w$ for each $\vec{x_s}_i \in D_S$, such that the owner can later obtain $Y_S$ by providing $D_S$ to $F_w$.
The sequence $D_S$ is referred to as \textit{watermark-carrier dataset} in the rest of our paper. 
Note that the watermark-carrier data distribution $p_s$ should be different from the training data distribution $p_x$. 
For example, one common way is to generate $D_S$ randomly as the $i$-th draw from a pseudo-random function.

\subsection{Existing Embedding Methods}
\label{sec:existing_methods}

Depending on the watermark representation, the watermark $S$ could be embedded in the model's parameters $w$, or the model's predictions $F_w(\vec{x})$ on the watermark-carrier set $D_S$.
Following the same spirit of generic watermarking techniques~\cite{cox1997secure}, the embedding should meet two requirements, i.e., \textit{fidelity} and \textit{robustness}. Fidelity requires the performance of the host network (i.e., the neural network to which the watermark bit vector is embedded into) 
is not impaired by the embedding, while robustness requires the embedded watermark to be detectable even if the host network undergoes modifications. 

\subsubsection{Embedding in Parameters $w$}
Essentially, this problem is to embed a $n$-bit vector $S \in \{0,1\}^n$ into $w$ of a given neural network (host network). 
The avenues where existing methods embed the watermark into $w$ typically falls into 4 classes: the least significant bit(s) of $w$, the signs of $w$, correlation with $w$ and statistics of $w$.

\underline{\textit{Least significant bit(s) of $w$} (W:LSB).}
Leveraging an observation that high-precision parameters are not necessary for high performance of the model~\cite{pruning_quantization}, Song et al. \cite{ccs_ML_remembers} investigated embedding the secret bit vector directly into the least significant bit(s) of the network parameters. They show that given a CNN model comprising $880$K parameters and  trained on the LFW dataset \cite{LFWTech}, the adversary can embed up to $17.6$M bits at a cost of $0.14\%$ decrease in test accuracy of the model. 

\underline{\textit{Signs of $w$} (W:SGN).} 
Another avenue that the model owner can exploit to embed his watermark is the signs of the model parameters \cite{ccs_ML_remembers}. In particular, given a watermark bit vector $S \in \{-1,1\}^n$, the owner would like to force the sign of $w_i$ to match that of $S_i$. The owner can achieve this by adding a penalty term $P$ to the original loss function. $P$ is defined as:
\begin{equation}
P (w, S) = \frac{\lambda_S}{n} \sum_{i=1}^n |\max(0, -w_i S_i)|
\end{equation}
where $\lambda_S$ controls the magnitude of the penalty. The penalty is minimal (i.e., zero) when $w_i$ and $S_i$ have the same sign.

\underline{\textit{Correlation with $w$} (W:COR).} 
Alternatively, the model owner can embed the watermark $S \in \mathbb{R}^l$ into the model parameters $w$ by adding a correlation term $C$ to the loss function that is employed during training~\cite{ccs_ML_remembers}, 
so as to maximize the correlation between $w$ and $S$. The correlation term $C (w, S)$ is defined as:
\begin{equation}
C(w, S) = - \lambda_c \cdot \frac{|\sum_{i=1}^l (w_i - \bar{w}) (S_i - \bar{S})|}{\sqrt{\sum_{i=1}^l (w_i - \bar{w})^2}\sqrt{\sum_{i=1}^l (S_i - \bar{S})^2}}
\end{equation}
where $l$ is number of parameters, $\lambda_c$ is the level of correlation and $\bar{w}, \bar{S}$ are mean values of $w, S$, respectively.

\underline{\textit{Statistics of $w$} (W:STA).} 
Uchida et al. \cite{embedding_watermarks} have investigated a problem of embedding watermarks into the statistical information of $w$ by the use of a regularization term $E_R(w, S)$ defined in the following. 
\begin{equation}
\label{eq:icmr}
E_R(w, S) = - \sum_{j=1}^n (S_j \log (y_j) + (1 - S_j) \log(1-y_j))
\end{equation}
where $y_j = \sigma (\sum_iX_{ji}w_i)$ and $\sigma(x) = \frac{1}{1 + \exp(-x)}$. The matrix $X$ is an embedding parameter (secret key) with size $n\times M$, where $M$ is the size of the network parameters $w$.
The regularization term enforces $w$ to have a certain statistical bias reflecting the embedded watermark. 
Experimental studies show that the watermark embedding incurs minimal effect on the performance of the host network (e.g., increase a test error rate on CIFAR-10 dataset by only $1\%$~\cite{embedding_watermarks}). They also show that the watermark remains detectable in the event the host network undergoes fine-tuning and compression by pruning. Nevertheless, we show later in our evaluation that the embedded watermark is detached if the host network is distilled. 
Besides, the parameter size $M$ in a neural network is usually numerous. Thus, the embedding parameter $X$ (with size $n\times M$) consumes massive memory especially when the watermark size $n$ is large. In practice, the amount of watermark embedded by this approach is quite limited.

Similarly, Chen et al.~\cite{deepmarks} and Rouhani et al.\cite{deepsigns} also add a regularization term to embed a watermark in the probability density function of the model's parameters or neurons' activations.

\subsubsection{Embedding in Predictions $F_w(\vec{x})$}

\underline{(P:CAP).}
The capability of neural networks to ``memorize'' random noise~\cite{rethinking_generalisation_DL} suggests that the model owner can force the model to ``memorize'' the watermark of his choice, and then rely on the model's predictions $F_w(\vec{x})$ to extract the watermark \cite{ccs_ML_remembers}.  
In particular, the owner synthesizes a set of records, and assigns to them labels $Y_S$ that encode the watermark he wants to embed, obtaining a labeled watermark-carrier set $D_S$. He then poisons the training data set $D$ with the synthetic data set $D_S$. Finally, he trains the model on the poisoned training set using a standard training pipeline. 
When $F_w$ becomes overfitted on $D_S$, the owner can extract the watermark from model predictions $F_w(\vec{x})$ by querying $F_w$ with $D_S$. 
Experimental results show that the amount of embedded information is equivalent to the pixel information of $25$ images in the CIFAR-10 dataset at a cost of $0.69\%$ decrease in the test accuracy of the model~\cite{ccs_ML_remembers}. 

Similarly, Adi et al.~\cite{watermark_backdoor} watermark a neural network by re-training it on a random set $D_S$ with random labels $Y_S$ as the watermark.
Merrer et al.~\cite{adversarial_watermark} use a set of adversarial examples of a neural network to convey the watermark information via a set of queries of them.

\subsection{Watermark Removal Attacks: Model Transformations}
\label{sec:transformations}

Given a trained model, transformation techniques aim at deriving a new model with the same or slightly different prediction task with additional features such as memory and computational efficiency, for example when the model is going to be used in mobile applications.  It might also be needed for transfer learning, updating the model using fine tuning, or further regularizing the model.  
Many of the transformation techniques make use of an extra refining set $D'$ in connection with the original model $F_w$ to update or reconstruct it as $F_w'$.
The model transformations can be used as attacks to remove the embedded watermarks in neural networks \cite{embedding_watermarks, watermark_backdoor, adversarial_watermark, deepsigns, deepmarks}.
Although there are many transformation techniques proposed in the literature, in this section, we only list the major techniques for deep neural networks, which are commonly used in practice.

\textit{\underline{Model Compression.}} 
The objective here is to optimize the memory needed to fit the (parameters of the) model, while preserving the accuracy of the model.  The compression can be achieved by removing insignificant parameters and pruning their links between neurons~\cite{han_pruning} (which is often followed by fine tuning the remaining parameters using the refining set to further make use of them), limiting the number of required bits to represent the model's parameters~\cite{pruning_quantization}, or grouping parameters into a few hash buckets~\cite{compress_by_hashing}. 

\textit{\underline{Distillation}}~\cite{distilling_DL} is another type of compression, where the original model's knowledge could be distilled into another model of smaller size, for example by reducing the number of neurons in each layer.  
Essentially, the knowledge of the original model $F_{w,T}$ (teacher model) is represented as its predictions on a refining dataset $D'$ (which is drawn from $p_x$) under temperature $T$ in the $\mathsf{softmax}$ function (see Section~\ref{sec:ml}). The temperature $T$ is usually set larger than $1$ so as to make the teacher model $F_{w,T}$ produce a softer prediction (i.e., softer probability distribution over classes), which encodes more knowledge of $F_{w,T}$.
The smaller model $F_w'$ (student model) is then trained on $D'$ with combination of the soft predictions produced by $F_{w,T}$ and hard labels which are determined by the ground truth labels of $D'$.

\textit{\underline{Fine Tuning.}}
A very common practice in machine learning is to refine and update a model using new data.  In the fine tuning process, an existing model $F_w$ is updated by simply training a new model $F_w'$ on a refining set $D'$, while the initial parameters of $F_w'$ are set to those of $F_w$~\cite{fine_tune}.

\textit{\underline{Transfer Learning.}}
This transformation technique is used to update the classification task of a model $F_w$ to a related yet slightly different task~\cite{transfer_learning_cvpr, transfer_learning_survey}.  
It often retains the lower layers of the original model, which usually extract generic features, and fine-tunes or retrains the last few layers using the refining set $D'$ as the training set for the new model $F_w'$.  

\textit{\underline{Computation Optimization.}}
The computation time for predictions on a test input is often not negligible for (deep) convolutional neural networks.  Using a technique known as {\em low-rank expansion}, it has been shown that approximating the convolutional layers of a model by linear combinations of smaller filters can accelerate the network's computation~\cite{expand_nonlinear_CNN, speedingNN_expansion}. In other words, a convolutional layer in a pre-trained model is decomposed into more convolutional layers with smaller filters, thus reducing the computation complexity~\cite{speedingNN_expansion}.  

\subsection{Robustness of Existing Embedding Methods}

\begin{table}[]
\centering
\setlength{\tabcolsep}{1.5pt}
\caption{Robustness of existing and our (P:ING) embedding methods. 
The accuracies of the main classification task and the watermark extraction are presented in white and gray columns respectively.
The host network is a fully-connected multilayer perceptron (MLP) trained on MNIST dataset. The size of watermark-carrier set $D_S$ is 1,000 images. We perform removal attacks on the embedded network including distillation (with temperature 5), pruning (with rate 0.4), rounding (by 2 digits) and fine tuning.
}
\label{tb:robustness_mnist}
\begin{tabular}{l|la|la|la|la|la}
\hline
 & \multicolumn{2}{c|}{\multirow{2}{*}{Embed}} & \multicolumn{8}{c}{Attack} \\
\cline{4-11} 
& \multicolumn{2}{c|}{} & \multicolumn{2}{c|}{Distillation} & \multicolumn{2}{c|}{Pruning} & \multicolumn{2}{c|}{Rounding} & \multicolumn{2}{c}{Fine Tuning} \\
\hline
W:LSB & 0.98 & 1.00 & 0.98 & 0 & 0.97 & 0.24 & 0.97 & 0 & 0.99 & 0.27 \\
W:SGN & 0.98 & 1.00 & 0.98 & 0 & 0.98 & 0.53 & 0.98 & 0.75 & 0.99 & 0.59 \\
W:COR & 0.98 & 0.99 & 0.98 & 0 & 0.97 & 0.42 & 0.97 & 0.41 & 0.99 & 0.48 \\
W:STA & 0.98 & 0.80 & 0.98 & 0 & 0.98 & 0.77 & 0.99 & 0.80 & 0.98 & 0.80 \\
P:CAP & 0.98 & 1.00 & 0.98 & 0 & 0.98 & 1.00 & 0.98 & 1.00 & 0.98 & 1.00 \\
\hline
\textbf{P:ING} & 0.97 & \textbf{1.00} & 0.97 & \textbf{0.23} & 0.98 & \textbf{1.00} & 0.97 & \textbf{1.00} & 0.97 & \textbf{1.00} \\ 
\hline
\end{tabular}
\end{table}

We first present part of our experimental results on the robustness of existing embedding methods in Table \ref{tb:robustness_mnist}.
The result shows that distillation is a strong attack which can remove all the embedded watermarks by existing methods. Nonetheless, our method (P:ING) preserves much more watermark information after distillation. Besides, it also achieves comparable robustness to the best result of existing methods against other attacks.

Embedding $S$ into parameters $w$ (i.e., W:$\ast$ methods) is inherently fragile in attacks that alter $w$ or the model architecture. For example, parameter pruning and rounding for compression purpose can completely remove watermarks embedded by W:LSB method. 
Distilling the model's knowledge to another model of smaller size from scratch destroys all the watermarks because it has a fresh model architecture and training process.

\begin{figure}[t]
\begin{center}
\includegraphics[width=70mm]{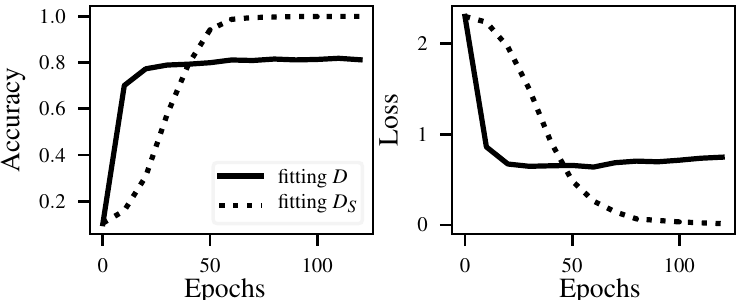}
\caption{Trajectories of training metrics on main task data $D$ and noise data $D_S$.
$D$ is the test set of CIFAR10.
The size of $D_S$ is $1,000$ random images which have random labels.
}
\label{fig:host_task_vs_noise}
\end{center}
\end{figure}

In existing methods that embed the watermarks into the model's predictions (i.e., P:CAP method),
the watermark-carrier set $D_S$ is synthesized randomly with labels on the owner's will, so it is \textit{noise data} to the model in terms of the main classification task.
A series of recent studies has indicated that
a model fits the noise data by ``memorization'' instead of extracting general patterns from it, and learns the general patterns of the real data first before fitting noise data~\cite{memorizattion_DL}. 
Our experiments align with this phenomenon as shown in Figure~\ref{fig:host_task_vs_noise}. 
The noise data $D_S$ is memorized in a later stage during training after the model learns the most meaningful features on $D$. 
This suggests that the embedded watermarks form a hidden content (as a set of parameters and neural connections) that is largely {\em independent} from the main classification task.  
In other words, there is a negligible knowledge intersection between the main classification task and the representation of watermarks in the model. 
Exactly because of this independence, the hidden watermark functions are very fragile to distillation whose major objective or constraint is to preserve the accuracy of the original model. Hence, redundant and independent embedded watermarks can be easily removed. 

\section{Ingrain}

\begin{figure}[t]
\begin{center}
\includegraphics[width=.85\linewidth]{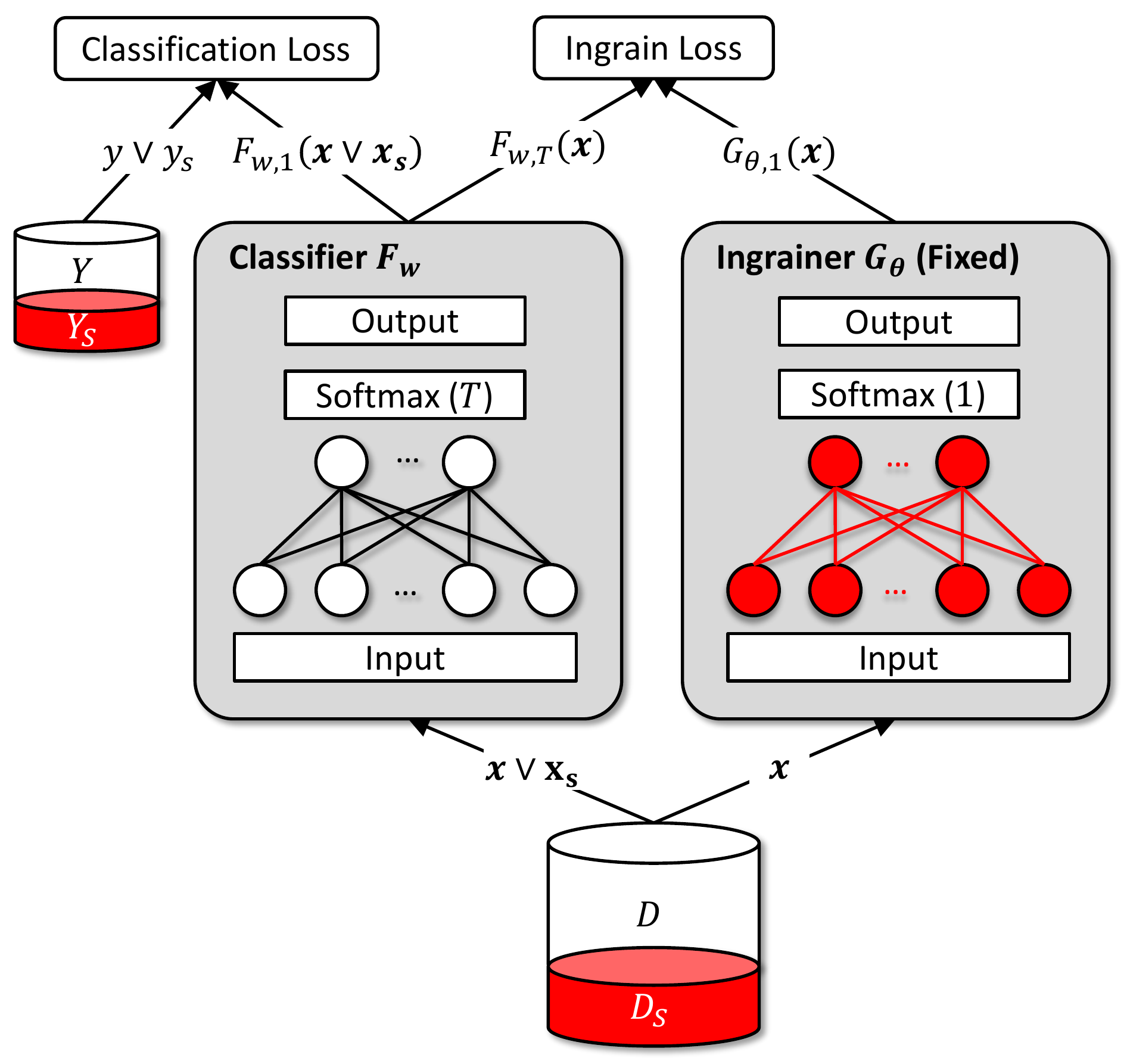}
\caption{Watermark ingrain in a neural network classifier.  The goal is to train a classifier $F_w$ that also carries the watermarks represented by an ingrainer model $G_\theta$. $G_\theta$ has the same input-output format and architecture as $F_w$, and is pre-trained on a watermark-carrier set $D_S$ (which together with its label sequence $Y_S$ compose the watermarks), so its parameters $\theta$ are fixed during the ingrain.  
During the ingrain, $F_w$'s training set $D$ is augmented with $D_S$ for the classification loss $\mathcal{L}(F_w(\mathbf{x}), y)$ to reinforce the embedding of watermarks in $F_w$.  
We use stochastic gradient descent (SGD) to update $F_w$'s parameters $w$, by jointly optimizing the classification loss function and the ingrain loss function $\mathcal{L}(F_{w,T}(\mathbf{x}), G_\theta(\mathbf{x}))$ weighted by an ingrain coefficient $\lambda$.  The ingrain loss acts similarly as a regularizer and helps $F_w$'s predictions on $D$ implicitly contain the watermark information.}
\label{fig:architecture}
\end{center}
\end{figure}

Drawing insights from the fragility of existing embedding methods especially to distillation, 
We introduce \textit{ingrain} as a means to deeply embed the watermarks into the model's functionalities by mitigating the independence between the watermark function and the main classification function.
We first present an overview of ingrain technique. Next, we introduce the ingrainer model which represents the watermarks.
We then use it to regularize the training of the classifier to embed the watermarks.

\subsection{Overview} 

In ingrain, the watermark $S$ is represented as the watermark-carrier sequence $D_S$ and its label sequence $Y_S$ in the embedding process.
Figure~\ref{fig:architecture} illustrates the ingrain mechanism, which is an indirect way of embedding watermarks in a neural network as opposed to directly overfitting the model on the watermarks.  The main idea behind ingrain is to force the model to carry the watermark information on its predictions on in-distribution data (i.e., data sampled from the training data distribution $p_x$).  Thus, when the model is attacked (i.e., distilled by using a refining dataset $D'$ drawn from $p_x$), the watermark knowledge is also transferred along with the model's core knowledge on classifying $D'$.  

The main technique of ingrain is to explicitly represent the watermark $S$ using an additional model---ingrainer, and further ingrain the watermark information implicitly in the classifier's predictions on training data $D$.
Such watermark information is expected to also appear in the classifier's predictions on the refining dataset $D'$ which is used in the further distillation attack.
This is achieved by modifying its training process. 
In particular, the ingrainer $G_\theta$, where $\theta$ are its parameters, can cast the watermark information to its predictions implicitly on $D$. When training the classifier, such ingrainer's predictions work as an additional term in the loss function, so as to encourage the classifier to \textit{simultaneously} learn the ground-truth labels and the watermark information (in ingrainer's predictions) on the \textit{same} training data $D$. Hence, the watermark information is correlated with the classifier's predictions on $D$ in an implicit way. 
Note that this does not mean the model owner extracts the explicit watermark $S$ using in-distribution inputs (i.e., training data $D$). 
It is actually the tricky part in the design of ingrainer where its predictions on training set $D$ carries the watermark information, but the owner extracts the watermark $S$ by querying the classifier with a watermark-carrier set $D_S$ which is drawn from a different distribution $p_s$.

\subsection{Ingrainer Model}
\label{sec:infuser}

The ingrainer model $G_\theta$ is used to represent the watermark information via its predictions on the training data $D$, such that it can regularize the training of the classifier to ingrain watermark information implicitly in the classifier's predictions on training data $D$.

The most straightforward way one might think of to fulfill such mapping is to train the ingrainer model directly on a subsequence of $D$ labeled by a sequence $Y_S$ which jointly compose the watermark $S$. 
However, it has several issues. Let $D_{sub}$ represent the chosen subsequence. On the one hand, if $D_{sub}$ is chosen in the way that 
the ground-truth label of each data point $\vec{x}_i \in D_{sub}$ is identical to ${y_s}_i\in Y_S$, it is exactly the classification task and does not watermark the model.
On the other hand, if $\vec{x}_i$'s ground-truth label is not ${y_s}_i$ and it is embedded as a hidden function, it watermarks the model but can be easily removed in attacks (e.g., distillation) that re-train the model on a refining set.

We sample the watermark-carrier set $D_S$ from a different distribution $p_s$ to carry $Y_S$.
We train the ingrainer, using the same architecture as the classifier, to overfit on $D_S$ (see Algorithm~\ref{alg:train_infuser}). Although this does not establish an explicit mapping from training data $D$ to the watermark $S$, it leads a part of the ingrainer's connections or parameters to memorize the watermark information. When a training data passes through the ingrainer, it is expected to trigger some of these connections, leading the output (prediction) to encode some implicit watermark information. The same architecture as the classifier is expected to boost this implicit mapping. Thus, the entire training set $D$ is supposed to carry rich watermark information in the ingrainer's predictions.
Although our experimental results demonstrate the effectiveness of the ingrainer by training it in this way, we believe it is not the only way. We leave an investigation into various potentially effective ingrainers as future work.

\begin{algorithm}
\small
\caption{Training Ingrainer $G_\theta$}\label{alg:train_infuser}
\SetKwInOut{Input}{Input}
\SetKwInOut{Output}{Output}
\Input{Watermark-carrier set $D_S$, number of epochs $P$, learning rate $\eta $, size of mini-batch $q$.}
\Output{Model parameters $\theta$ of ingrainer $G_\theta$.}

$\theta \gets $ \textbf{initialize}($G_\theta$) \\
\For{\text{p = 1 to P}}{
	
	\For{\text{each mini-batch $\{(\vec{x_s}_j, {y_s}_j)\}_{j=1}^q \subset D_S$}}{
		$g \gets \nabla_\theta\frac{1}{q}\sum_{j=1}^q    {\mathcal L} (G_\theta(\vec{x_s}_j), {y_s}_j)$ \\
		$\theta \gets$ \textbf{updateParameters} $(\eta, \theta, g)$
	}
	
}
\end{algorithm}

\subsection{Training Classifier}

The training of the classifier $F_w$ is regularized by the trained ingrainer model $G_\theta$ to force $F_w$ to learn two tasks together. 
Specifically, we add an additional term $\mathcal{L}(F_{w,T}(\mathbf{x}), G_\theta(\mathbf{x}))$ to the training loss function $\mathcal{L}(F_w(\mathbf{x}), y)$, where $T$ determines the classifier's temperature in the $\mathsf{softmax}$ function.
This term is referred to as \textit{ingrain loss} in the paper.

In the distillation attack, the classifier $F_{w,T}$ becomes the teacher model whose $T$ is usually set larger than $1$ to produce soft predictions which also encode the watermark knowledge. The student model learns the $F_{w,T}$'s knowledge from the soft predictions.
To maximally preserve  watermark information in such soft predictions, we set a larger $T$ for $F_{w,T}$ in the ingrain loss to ingrain $G_\theta$'s watermark information in $F_{w,T}$'s soft predictions.
It mitigates the distance between the soft predictions $F_{w,T}(\mathbf{x})$ on each training data $\vec{x}$ and the watermark information $G_\theta(\mathbf{x})$ carried by $\vec{x}$ when passing through $G_\theta$. 
This leads to a correlation of the classifier's soft predictions with the watermark information, which improves the watermark's resistance against the distillation attack. 
Note that the ingrain process is similar to distillation where $G_\theta$ is the teacher model and $F_{w,T}$ is the student model. Nonetheless, we set a higher $T$ for the student model as opposed to the teacher model.
Formally, the loss function we are optimizing when training the classifier is the following.
\begin{align}
\label{eq:loss}
\begin{split}
 L_D(F_w) =  \frac{1}{|D|} \sum_{\vec{x}\in D} & \mathcal{L}(F_w(\mathbf{x}), y) \\
 & + \lambda \mathcal{L}(F_{w,T}(\mathbf{x}), G_\theta(\mathbf{x}))
\end{split}
\end{align}
where the ingrain coefficient $\lambda$ determines the degree of ingrain. 

Throughout the training process, the ingrainer is fixed, and we use SGD (see Section~\ref{sec:train}) to optimize the classifier's parameters $w$. 
Each training data, that passes through the ingrainer, produces one sample point in the watermark space. The classifier that is regularized with these samples, tries to also solve for the watermark function through the optimization process.  Thus, by tuning the ingrain coefficient $\lambda$, we can control the trade-off between the prediction accuracy of the classifier and the degree to which the watermark information is ingrained in the classifier's predictions on training data.

\underline{{\em Explicitly embed watermarks into classifier.}} 
The ingrain loss imbues the classifier's predictions on training data with implicit watermark information, which can lead to better robustness against distillation attack. Nonetheless, it does not explicitly embed the watermarks into the classifier. Querying $F_w$ with watermark-carrier set $D_S$ might not obtain the $Y_S$ with high accuracy.
Therefore, we augment (poison) the training data $D$ with $D_S$ in the training of $F_w$ using loss function $\mathcal{L}(F_w(\mathbf{x}), y)$.
This is to further enforce the embedding into the unused parameter space of the model. In this case, when the ingrain coefficient $\lambda$ is set to 0, the whole process becomes equivalent to the existing P:CAP method that exploit the neural networks' large capacity~\cite{ccs_ML_remembers}. 
The watermark information that is passed to the classifier through poisoning $D$ with $D_S$ is not expected to be resistant to distillation attack, but it helps boosting the accuracy of the watermark extraction for the case of simple attacks, e.g., parameter pruning, rounding.

\begin{figure}[t]
\centering   
\includegraphics[width=0.7\linewidth]{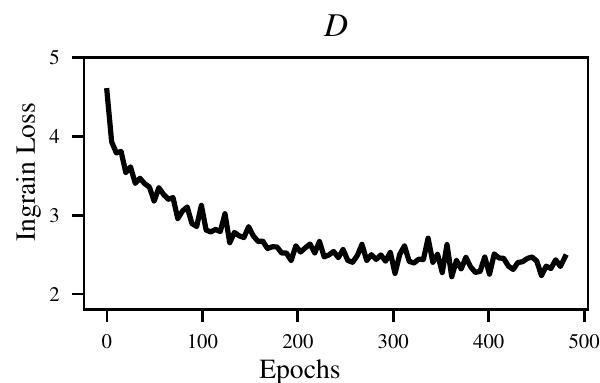}

\includegraphics[width=0.7\linewidth]{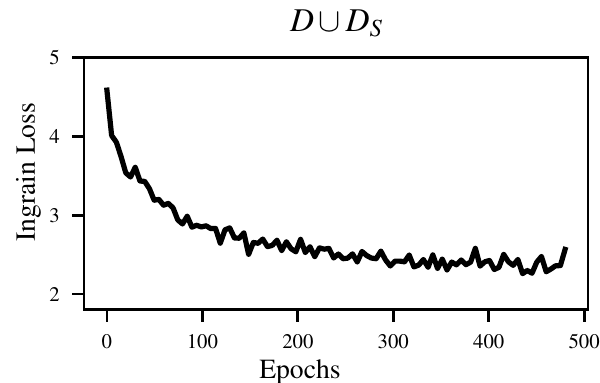}
\caption{
Ingrain loss on the training data $D$ in the training of the classifier. (Top) $D$ is not augmented. (Bottom) We use $D_S$ to augment $D$ in the classification loss.
The host network is a MLP model. $D$ is MNIST dataset. The size of $D_S$ is $1,000$ images. The ingrain temperature $T$ is $10$. The ingrain coefficient $\lambda$ is $2$.}\label{fig:loss}
\end{figure}

A further interesting finding is that such augmentation imposes negligible influence on the ingrain loss as shown in Figure \ref{fig:loss}.
This result demonstrates the independence of the sub-models that are responsible for memorizing the watermarks and the part of the model which represents the classification task, as in the case of existing P:CAP method~\cite{ccs_ML_remembers}. 
Such independence leads to the fragility of P:CAP against distillation attack.
In summary, the whole process of training the classifier is shown in Algorithm~\ref{alg:infusion}, where we shuffle $D$ and $D_S$ (line 2), and extract them in each mini-batch (line 5-6) to apply separate losses on them (line 7-9). The weighted average gradient is used to update the parameters $w$ (line 10-11).

\begin{algorithm}
\small
\caption{Training Classifier $F_w$}\label{alg:infusion}
\SetKwInOut{Input}{Input}
\SetKwInOut{Output}{Output}

\Input{Training dataset $D=\{(\vec{x}_j,y_j)\}_{i=1}^n$, watermark-carrier set $D_S=\{(\vec{x_s}_j,{y_s}_j)\}_{i=1}^m$, Ingrainer $G_\theta$, number of epochs $P$, learning rate $\eta $, ingrain coefficient $\lambda$, ingrain temperature $T$.}
\Output{Model parameters $w$ of classifier $F_w$.}

$w \gets $ \textbf{initialize}($F_w$) \\
$D_A \gets$ \textbf{shuffle}($D \cup D_S$) \\

\For{\text{p = 1 to P}}{
	\For{\text{each mini-batch $b\subset D_A$}} {
		$\{(\vec{x}_j, y_j)\}_{j=1}^a \gets$ \textbf{getTrainData}($b$) \\
		$\{(\vec{x_s}_j, {y_s}_j)\}_{j=1}^b \gets$ \textbf{getWatermarkCarrier}($b$) \\
		$g_D \gets \nabla_w\frac{1}{a}\sum_{j=1}^a    {\mathcal L} (F_w(\vec{x}_j), y_j) $\\\hspace{2.3cm}$ + \lambda {\mathcal L} ( F_{w,T}(\vec{x}_j), G_\theta(\vec{x})) $ \\
		$g_{D_S} \gets \nabla_w\frac{1}{b}\sum_{j=1}^b    {\mathcal L} (F_w(\vec{x_s}_j), {y_s}_j)$ \\
		$g \gets (a  g_D+b  g_{D_S})/(a+b)$ \\
		$w \gets$ \textbf{updateParameters} $(\eta, w, g)$ \\
	}
}
\end{algorithm}

\section{Experiments}

In this section, we perform an empirical comparison of existing watermark embedding methods and ingrain. 
We first introduce the datasets and models we used, and the embedding methods and removal techniques we evaluated. 
Then, we compare them in both embedding performance and embedding robustness.

\subsection{Datasets \& Models}
We perform the evaluation on benchmark image datasets including CIFAR10~\cite{krizhevsky2014cifar} and MNIST~\cite{lecun1998mnist}. 
We train convolutional neural network (CNN) and fully-connected multi-layer perceptron (MLP) for these datasets.
\begin{itemize}[leftmargin=2em]
\item \textbf{MNIST~\cite{lecun1998mnist}.} This is a dataset composed of 70,000 handwritten digit images with size 28$\times$28 in 10 classes. We rescale each image to value range [0, 1]. 
\textbf{Model}. We use a fully-connected neural network with the same architecture as in ~\cite{distilling_DL}. Specifically, it has two hidden layers of 1200 rectified linear neurons. Each fully-connected layer is regularized using dropout 0.5. We train the model in mini-batch size 128, using the Adadelta~\cite{zeiler2012adadelta} optimizer with initial learning rate 0.1, $\rho$ 0.95, and $\varepsilon$ 1e-8. 

\item \textbf{CIFAR10~\cite{krizhevsky2014cifar}.} It consists of 32$\times$32 color images in 10 classes. It has 50,000 training records and 10,000 test records. We rescale each image to value range [0, 1]. \textbf{Model}. We train a standard convolutional neural network (CNN) with the same architecture as in ~\cite{papernot2016distillation}. Specifically, it is a succession of 2 convolutional layers with 64 (3$\times$3) filters, a max pooling layer, 2 convolutional layers with 128 (3$\times$3) filters, a max pooling layer, and 2 fully connected layers with 256 neurons. Each fully-connected layer is regularized with dropout 0.5. We set the mini-batch size to 128.
We use SGD with momentum 0.9 to optimize the model. The initial learning rate is set to 0.01, and decays 0.95 every 10 epochs. 
\end{itemize}

\subsection{Embedding Techniques}

We evaluate the following embedding techniques.

\begin{itemize}[leftmargin=2em]
\item \textbf{W:LSB}~\cite{ccs_ML_remembers}, embedding into the least significant bit(s) of parameters. Trained for 500 epochs.
\item \textbf{W:SGN}~\cite{ccs_ML_remembers}, embedding into the sign of parameters. Trained for 500 epochs.
\item \textbf{W:COR}~\cite{ccs_ML_remembers}, embedding into parameters by making them correlated to the watermarks. Trained for 500 epochs.
\item \textbf{W:STA}~\cite{embedding_watermarks}, embedding into the statistical information of parameters. Trained for 500 epochs. 
\item \textbf{P:CAP}~\cite{ccs_ML_remembers,watermark_backdoor}, embedding into predictions on a watermark-carrier set through capacity abuse. Trained for 500 epochs.
\item  \textbf{P:ING}, embedding into predictions through ingrain. We train the ingrainer for 1,000 epochs, and train the classifier for 500 epochs. We perform ingrain 5 times with ingrain coefficient $\lambda$ of 0.5, 1, 2, 4, and 8 respectively. We set the ingrain temperature $T$ to 10 during ingrain, because we empirically find it gives the best overall embedding robustness.
\end{itemize}

\noindent \textbf{Watermark-carrier set $D_S$}.
Similar to the setting of the P:CAP method~\cite{ccs_ML_remembers}, we compose $D_S$ by randomly generating images using a pseudorandom number generator based on a random seed.  Such data points could be generated in many different ways, for example, white noise, uniformly random noise, one-hot images~\cite{ccs_ML_remembers}, and random walk.
Random walk starts from the center of a blank image and moves by one pixel of random value at a random direction for many steps---which we set to the size of the image.  In our experiments, we observe that random-walk images are more effective in practice (i.e., easier to embed with less accuracy overhead). We use it to generate $D_S$. 
Totally, we have synthesized three $D_S$ sets with size 1,000, 5,000 and 10,000 watermark-carrier images respectively. Each $D_S$ set is labeled with a sequence $Y_S$ which jointly compose the watermarks.
For W:$*$ methods, we convert $D_S$ to a number of bits that contain the same amount of watermark information.
Note that W:STAT needs massive memory in the embedding process (see Section \ref{sec:existing_methods}), which makes it impractical to embed the same amount of watermark information. We embed 256 bits for it, as in the case of \cite{embedding_watermarks}.

\subsection{Watermark Removal Attacks}
We evaluate the following watermark removal attacks in our experiments.
\begin{itemize}[leftmargin=2em]
\item \textbf{Distillation}~\cite{distilling_DL}, using a refining set and distilled for 500 epochs under different temperatures, i.e., 5, 10, and 15.
\item \textbf{Pruning}~\cite{pruning_quantization}, by pruning the parameters with different rates (from 0.1 to 0.6) followed by fine tuning trained for 25 epochs.
\item \textbf{Rounding}~\cite{pruning_quantization}, by reducing the precision of the parameters by 1 to 6 digits.
\item \textbf{Fine tuning}~\cite{fine_tune}, by retraining the model on the refining set for 25 epochs.
\item \textbf{Expansion}~\cite{expand_nonlinear_CNN, speedingNN_expansion}, by expanding the convolutional layers to linear combinations of smaller filters, for speedup rate ranging from 2.5x to 3x. It is not applicable to MLP models (i.e., classifier for MNIST).
\end{itemize}

For distillation and fine tuning attacks, we leave $20\%$ of the original training data $D$ as the refining set $D'$.  We use the remaining $80\%$ training data for training the classifier. 
The student model architectures in distillation are presented in the following.
\begin{itemize}[leftmargin=2em]
\item \textbf{MNIST}. The student model is a fully-connected neural network with 2 hidden layers of 800 neurons, which is the same architecture as in ~\cite{distilling_DL}. 

\item \textbf{CIFAR10}. It is a standard convolutional neural network, a succession of 2 convolutional layers with 64 (3$\times$3) filters, a max pooling layer, a convolutional layer with 128 (3$\times$3) filters, a max pooling layer, and 2 fully-connected layers of 200 neurons.
\end{itemize}

\subsection{Evaluation Metrics}

Embedding watermarks should preserve the classification accuracy comparable to the clean model. We evaluate the performance of a model using two metrics: \textit{classification accuracy} and \textit{watermark accuracy}.

The classification accuracy reflects the model's main classification performance. It is measured on an independent test set.
The watermark accuracy is the accuracy in extracting the watermarks. 
In P:$*$ methods, it is measured by using the watermark-carrier set $D_S$ to query the classifier $F_w$ and counting the ratio of correct predictions to $Y_S$.
In W:$*$ methods, it is measured by extracting the embedded watermark bits from the model and calculating the accuracy.
To better compare the watermark accuracy among different classifiers embedded using different methods, we normalize it as in the following.
$$\hat{A}_{wm}=\max \left (\frac{A_{wm}-\frac{1}{c}}{1-\frac{1}{c}},0\right )$$
where $A_{wm}$ represents the watermark accuracy, $c$ is the number of values that the watermark can take (i.e., $c$ equals to the number of classes in P:$*$ methods and $c=2$ in W:$*$ methods), and $\frac{1}{c}$ is the probability of random guess.

\subsection{Embedding Performance}

\begin{table*}[t]
\centering
\setlength{\tabcolsep}{4pt}
\caption{Embedding performance. The classification and watermark accuracies are presented in white and gray columns respectively. Numbers in the 2nd row under P:ING represent the $\lambda$.}
\label{tb:app_embedding}
\begin{tabular}{ll|la|lalalalala|lalalalala}
\hline
\multicolumn{1}{c}{} & \multicolumn{1}{c}{} & \multicolumn{2}{c|}{} & \multicolumn{10}{c|}{Prior Methods} & \multicolumn{10}{c}{P:ING} \\ \hline
\multicolumn{1}{c}{$|D_S|$} & \multicolumn{1}{c}{$D$} & \multicolumn{2}{c|}{Clean} & \multicolumn{2}{c}{W:LSB} & \multicolumn{2}{c}{W:SGN} & \multicolumn{2}{c}{W:COR} & \multicolumn{2}{c}{W:STA} & \multicolumn{2}{c|}{P:CAP} & \multicolumn{2}{c}{0.5} & \multicolumn{2}{c}{1} & \multicolumn{2}{c}{2} & \multicolumn{2}{c}{4} & \multicolumn{2}{c}{8} \\ \hline
1,000 & MNIST & 0.98 & 0.02 & 0.98 & 1 & 0.98 & 1 & 0.98 & 0.99 & 0.98 & 0.8 & 0.98 & 1 & 0.97 & 1 & 0.96 & 1 & 0.95 & 1 & 0.95 & 1 & 0.93 & 1 \\
& CIFAR10 & 0.84 & 0 & 0.81 & 1 & 0.81 & 0.99 & 0.8 & 0.98 & 0.8 & 0.98 & 0.83 & 1 & 0.82 & 1 & 0.81 & 1 & 0.79 & 1 & 0.76 & 1 & 0.71 & 1 \\
\hline
5,000 & MNIST & 0.98 & 0 & 0.98 & 1 & 0.97 & 1 & 0.98 & 0.98 & 0.98 & 0.8 & 0.98 & 1 & 0.96 & 1 & 0.96 & 1 & 0.95 & 1 & 0.94 & 1 & 0.92 & 1 \\
& CIFAR10 & 0.84 & 0 & 0.8 & 1 & 0.8 & 0.99 & 0.81 & 0.98 & 0.8 & 0.98 & 0.83 & 1 & 0.82 & 1 & 0.82 & 1 & 0.81 & 1 & 0.78 & 1 & 0.73 & 1 \\ 
\hline
10,000 & MNIST & 0.98 & 0 & 0.98 & 1 & 0.99 & 1 & 0.98 & 0.98 & 0.98 & 0.8 & 0.98 & 1 & 0.97 & 1 & 0.96 & 1 & 0.95 & 1 & 0.94 & 1 & 0.91 & 1 \\
& CIFAR10 & 0.84 & 0 & 0.79 & 1 & 0.8 & 0.99 & 0.8 & 0.98 & 0.8 & 0.98 & 0.83 & 1 & 0.82 & 1 & 0.82 & 1 & 0.81 & 1 & 0.79 & 1 & 0.76 & 1 \\ 
\hline
\end{tabular}
\end{table*}

\begin{figure*}[t]
\begin{center}
\includegraphics[width=\linewidth]{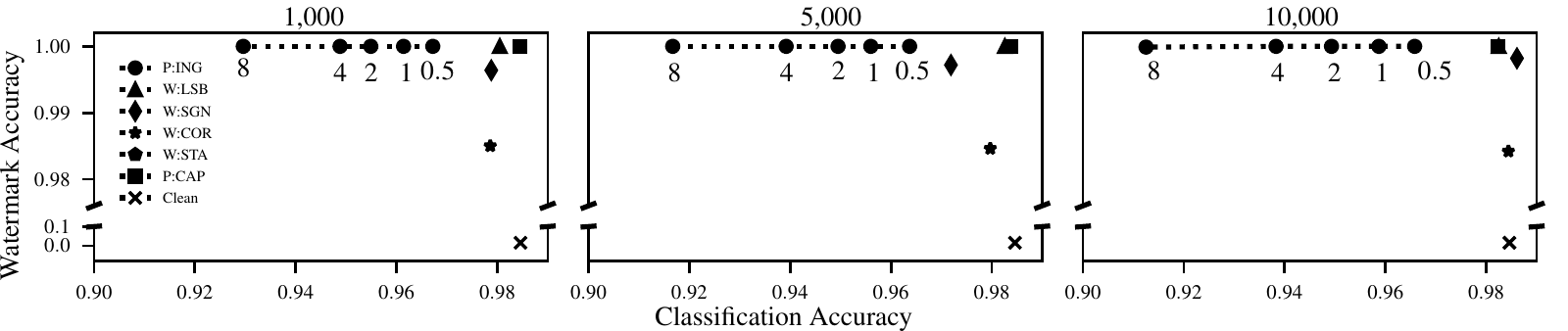}
\caption{Embedding performance. The host network is trained on MNIST. The size of $D_S$ is 1,000 (left), 5,000 (middle) and 10,000 (right) images. The ingrain coefficient $\lambda$s in P:ING are labeled as different numbers in the dashed curves.}
\label{fig:mnist_embed_all_wm_num}
\end{center}
\end{figure*}

\begin{figure*}[t]
\begin{center}
\includegraphics[width=\linewidth]{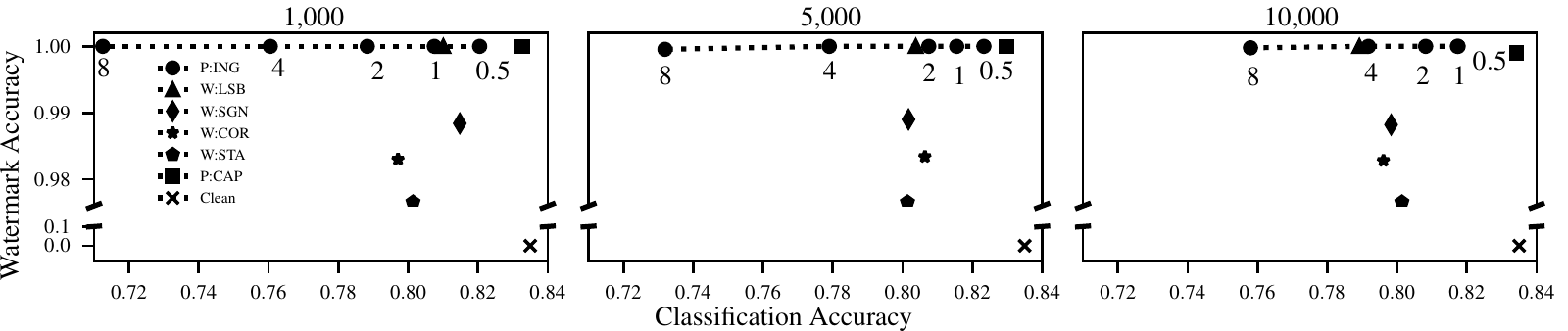}
\caption{Embedding performance. The host network is trained on CIFAR10. The size of $D_S$ is 1,000 (left), 5,000 (middle) and 10,000 (right) images. The ingrain coefficient $\lambda$s in P:ING are labeled as different numbers in the dashed curves.}
\label{fig:cifar10_embed_all_wm_num}
\end{center}
\end{figure*}

We depict the embedding performance of W:$*$ and P:$*$ methods in Figure \ref{fig:mnist_embed_all_wm_num} and \ref{fig:cifar10_embed_all_wm_num}.
The result shows that P:$*$ methods and W:$LSB$ achieves 100\% watermark accuracy, while the rest W:$*$ methods gets lower watermark accuracy.
It is worth noting that ingrain leads to a small drop (e.g., 1\%$\sim$5\% for MNIST) in the model's classification accuracy, which is proportional to the ingrain coefficient $\lambda$. 
This is because the $\lambda$ controls the weight of the classification loss and the ingrain loss during the training. A larger $\lambda$ leads to more watermark information ingrained in the model, but at the cost of losing more classification accuracy. It is a trade off between the two accuracies.
We present the comprehensive evaluation result of embedding performance in Table \ref{tb:app_embedding}.

\subsection{Embedding Robustness against Distillation}
\label{sec:robustness_to_distillation}

\begin{table*}[t]
\centering
\setlength{\tabcolsep}{4.5pt}
\caption{Embedding robustness against distillation. The classification and watermark accuracies are presented in white and gray columns respectively. Numbers in the 2nd row under P:ING represent the $\lambda$. The P:ING results that have better robustness than P:CAP are labeled in bold type.}
\label{tb:app_dist}
\begin{tabular}{ll|l|la|la|lalalalala}
\hline
\multicolumn{1}{c}{} & \multicolumn{1}{c}{} & \multicolumn{1}{c}{} & \multicolumn{2}{c|}{}  & \multicolumn{2}{c|}{Prior Method} & \multicolumn{10}{c}{P:ING} \\ \hline
\multicolumn{1}{c}{$|D_S|$} & \multicolumn{1}{c}{$D$} & \multicolumn{1}{c}{$T$} & \multicolumn{2}{c|}{Clean} & \multicolumn{2}{c|}{P:CAP} & \multicolumn{2}{c}{0.5} & \multicolumn{2}{c}{1} & \multicolumn{2}{c}{2} & \multicolumn{2}{c}{4} & \multicolumn{2}{c}{8} \\ \hline
&  & 5 & 0.984 & 0.018 & 0.984 & 0.036 & \textbf{0.97} & \textbf{0.229} & \textbf{0.965} & \textbf{0.271} & \textbf{0.96} & \textbf{0.297} & \textbf{0.958} & \textbf{0.339} & \textbf{0.955} & \textbf{0.34} \\
1,000 & MNIST & 10 & 0.982 & 0.02 & 0.983 & 0.037 & \textbf{0.969} & \textbf{0.224} & \textbf{0.964} & \textbf{0.261} & \textbf{0.96} & \textbf{0.288} & \textbf{0.957} & \textbf{0.307} & \textbf{0.956} & \textbf{0.327} \\
 &  & 15 & 0.982 & 0.019 & 0.982 & 0.032 & \textbf{0.969} & \textbf{0.197} & \textbf{0.966} & \textbf{0.242} & \textbf{0.962} & \textbf{0.262} & \textbf{0.96} & \textbf{0.279} & \textbf{0.959} & \textbf{0.281} \\ 
  \hhline{~|-|-|-|-|-|-|-|-|-|-|-|-|-|-|-|-|}
 &  & 5 & 0.817 & 0.006 & 0.816 & 0 & \textbf{0.79} & \textbf{0.009} & \textbf{0.772} & \textbf{0.088} & \textbf{0.755} & \textbf{0.13} & \textbf{0.739} &\textbf{ 0.151} & \textbf{0.724} & \textbf{0.181} \\
 & CIFAR10 & 10 & 0.81 & 0.008 & 0.809 & 0 & \textbf{0.795} & \textbf{0.033} & \textbf{0.775} & \textbf{0.053} & \textbf{0.759} & \textbf{0.101} & \textbf{0.752} & \textbf{0.116} & \textbf{0.741} & \textbf{0.147} \\
 &  & 15 & 0.803 & 0.002 & 0.802 & 0.002 & \textbf{0.794} & \textbf{0.03} & \textbf{0.784} & \textbf{0.06} & \textbf{0.769} & \textbf{0.07} & \textbf{0.764} & \textbf{0.103} & \textbf{0.763} & \textbf{0.101} \\ 
\hline
&  & 5 & 0.984 & 0 & 0.984 & 0.016 & \textbf{0.969} & \textbf{0.101} & \textbf{0.962} & \textbf{0.125} & \textbf{0.957} & \textbf{0.14} & \textbf{0.954} & \textbf{0.143} & \textbf{0.95} & \textbf{0.146} \\
5,000 & MNIST & 10 & 0.982 & 0 & 0.982 & 0.02 & \textbf{0.967} & \textbf{0.096} & \textbf{0.962} & \textbf{0.119} & \textbf{0.958} & \textbf{0.134} & \textbf{0.955} & \textbf{0.143} & \textbf{0.953} & \textbf{0.143} \\
 &  & 15 & 0.982 & 0 & 0.982 & 0.016 & \textbf{0.969} & \textbf{0.082} & \textbf{0.964} & \textbf{0.107} & \textbf{0.96} & \textbf{0.121} & \textbf{0.957} & \textbf{0.128} & \textbf{0.956} & \textbf{0.132} \\ 
  \hhline{~|-|-|-|-|-|-|-|-|-|-|-|-|-|-|-|-|}
 &  & 5 & 0.817 & 0 & 0.815 & 0 & 0.803 & 0 & \textbf{0.799} & \textbf{0.012} & \textbf{0.778} & \textbf{0.022} & \textbf{0.766} & \textbf{0.032} & \textbf{0.748} & \textbf{0.042} \\ 
 & CIFAR10 & 10 & 0.81 & 0 & 0.806 & 0 & \textbf{0.8} & \textbf{0.002} & \textbf{0.795} & \textbf{0.009} & \textbf{0.786} & \textbf{0.018} & \textbf{0.775} & \textbf{0.019} & \textbf{0.771} & \textbf{0.03} \\
 &  & 15 & 0.803 & 0 & 0.799 & 0 & 0.797 & 0 & \textbf{0.793} & \textbf{0.005} & \textbf{0.791} & \textbf{0.005} & \textbf{0.789} & \textbf{0.011} & \textbf{0.785} & \textbf{0.014} \\ 
\hline
&  & 5 & 0.984 & 0 & 0.984 & 0.005 & \textbf{0.971} & \textbf{0.054} & \textbf{0.966} & \textbf{0.065} & \textbf{0.959} & \textbf{0.077} & \textbf{0.956} & \textbf{0.083} & \textbf{0.952} & \textbf{0.087} \\
10,000 & MNIST & 10 & 0.982 & 0 & 0.982 & 0.007 & \textbf{0.969} & \textbf{0.047} & \textbf{0.965} & \textbf{0.061} & \textbf{0.961} & \textbf{0.073} & \textbf{0.957} & \textbf{0.083} & \textbf{0.955} & \textbf{0.083} \\
 &  & 15 & 0.982 & 0 & 0.981 & 0.006 & \textbf{0.971} & \textbf{0.037} & \textbf{0.966} & \textbf{0.052} & \textbf{0.963} & \textbf{0.066} & \textbf{0.961} & \textbf{0.071} & \textbf{0.958} & \textbf{0.072} \\ 
  \hhline{~|-|-|-|-|-|-|-|-|-|-|-|-|-|-|-|-|}
 &  & 5 & 0.817 & 0 & 0.815 & 0.004 & 0.802 & 0 & 0.796 & 0 & \textbf{0.797} & \textbf{0.007} & \textbf{0.782} & \textbf{0.009} & \textbf{0.77} & \textbf{0.018} \\
 & CIFAR10 & 10 & 0.81 & 0 & 0.81 & 0.001 & \textbf{0.794} & \textbf{0.002} & \textbf{0.795} & \textbf{0.002} & \textbf{0.792} & \textbf{0.003} & \textbf{0.791} & \textbf{0.006} & \textbf{0.787} & \textbf{0.008} \\
 &  & 15 & 0.803 & 0 & 0.804 & 0 & 0.791 & 0 & \textbf{0.794} & \textbf{0.01} & \textbf{0.793} & \textbf{0.005} & \textbf{0.792} & \textbf{0.005} &\textbf{0.793} & \textbf{0.007} \\ 
\hline
\end{tabular}
\end{table*}

The W:$*$ methods have no robustness against distillation because the original classifier is replaced by a new classifier with smaller size. All the original parameters in which the watermarks are embedded are destroyed.
Therefore, we only evaluate the robustness of P:$*$ methods.
We first present all the evaluation results on the robustness against distillation in Table \ref{tb:app_dist}.
The result shows that for each distillation temperature on each host classifier, there is at least one $\lambda$ that enables P:ING to embed watermarks with better robustness than P:CAP.

\begin{figure*}[t]
\begin{center}
\includegraphics[width=\linewidth]{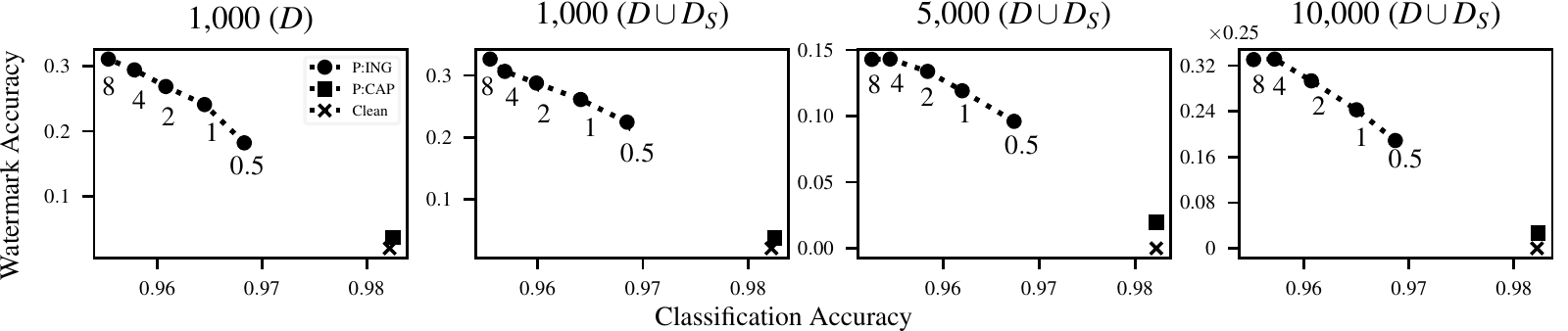}
\caption{Embedding robustness against distillation. The host network is trained on MNIST.
The size of $D_S$ is 1,000 (first 2 columns), 5,000 (3rd column), and 10,000 (4th column) images. 
The result in the 1st column is obtained without augmenting $D$ with $D_S$.
The $\lambda$s in P:ING are labeled as different numbers in the dashed curves. Distillation temperature is 10.}
\label{fig:mnist_dist_all_wm_num}
\end{center}
\end{figure*}

\begin{figure*}[t]
\begin{center}
\includegraphics[width=\linewidth]{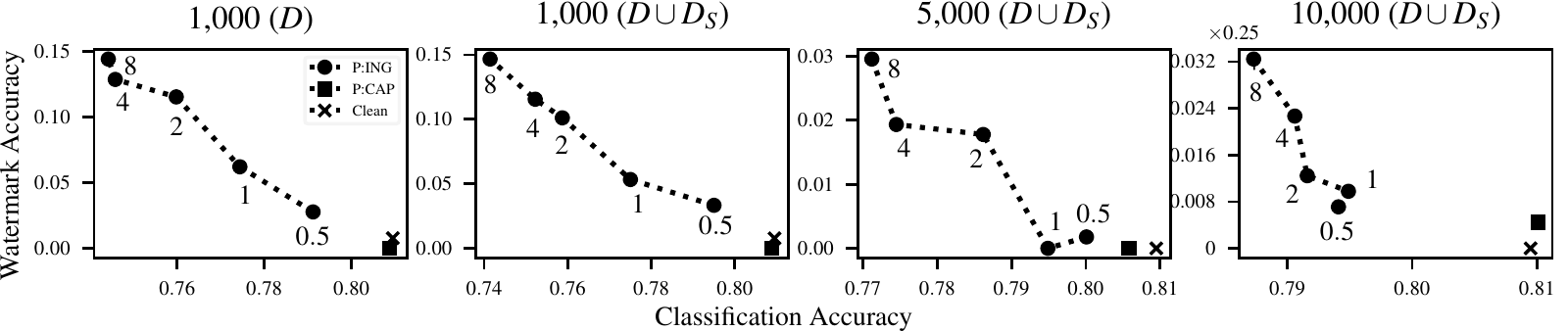}
\caption{Embedding robustness against distillation. The host network is trained on CIFAR10.
The size of $D_S$ is 1,000 (first 2 columns), 5,000 (3rd column), and 10,000 (4th column) images. 
The result in the 1st column is obtained without augmenting $D$ with $D_S$.
The $\lambda$s in P:ING are labeled as different numbers in the dashed curves. Distillation temperature is 10.}
\label{fig:cifar10_dist_all_wm_num}
\end{center}
\end{figure*}

\begin{figure}[t]
\centering    
\includegraphics[width=0.8\linewidth]{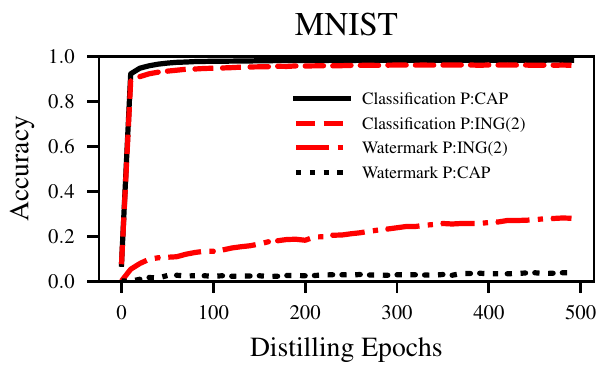}
\\
\includegraphics[width=0.8\linewidth]{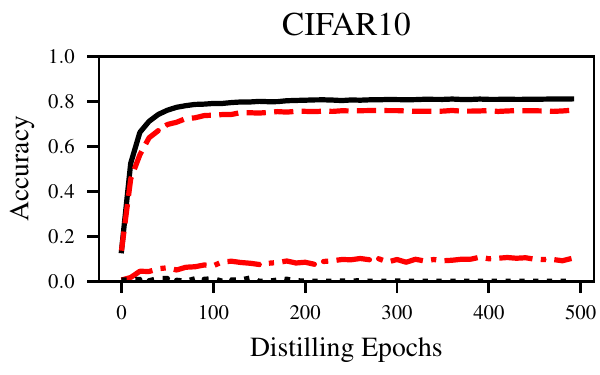}
\caption{Trajectories of classification and watermark accuracies during distillation. 
We compare P:CAP and P:ING wherein $\lambda=2$. Distillation temperature is 10.
}\label{fig:distillation_privacy}
\end{figure}

We depict the robustness of P:$*$ methods for MNIST and CIFAR10 classifiers in Figure \ref{fig:mnist_dist_all_wm_num} and \ref{fig:cifar10_dist_all_wm_num} under distillation $T=10$.
As it is shown, P:CAP has almost no robustness to distillation. On the other hand, P:ING preserves much more watermark information after distillation. For example, P:ING achieves 22.4\%$\sim$32.7\% watermark accuracy for MNIST classifier when $|D_S|=1,000$. As a trade off, the classification accuracy drops only 1.3\%$\sim$2.6\%.
As expected, the watermark accuracy (which reflects the embedding robustness) is proportional to $\lambda$.
We show the trajectories of the classification and watermark accuracies of P:CAP and P:ING during the distillation process in Figure \ref{fig:distillation_privacy}, where $\lambda=2$, distillation $T=10$.
The plots show that there is no significant difference between the classification accuracy of P:CAP and P:ING. 
However, as the model's distillation progresses, the ingrained model performs evidently better than the alternative approach in terms of the watermark accuracy.
This is mainly because the outputs of the classifier on the refining dataset $D'$ (i.e., the training set of distillation but are drawn from $p_x$ as well) also contain the watermark information when the watermark is ingrained in the classifier by P:ING.  Whereas, P:CAP embeds watermarks independently from the classifier's main task, which explains why the watermarks cannot be transferred to the student model during distillation. 

From the results, it is clear that a greater $\lambda$ leads to a more robust embedding, even on the 10,000 watermark-carrier data, where a $\lambda$ of 8 causes 1\% drop in embedding accuracy. Together, the present findings confirm that ingrain does help attain a more robust watermark embedding against distillation.  Figure~\ref{fig:distillation_privacy} compares the classification and embedding accuracies of the two embedding techniques (P:CAP and ingrain P:ING(2) with coefficient $\lambda=2$) during distillation. As the plots show, there is no significant difference between the classification accuracy of P:CAP and P:ING(2).  However, as the model's distillation progresses, the ingrained model performs evidently better than the alternative approach in terms of the embedding accuracy. This is mainly because the outputs of the model on benign data contain watermark information, when the watermark is embedded using ingrain.  Whereas, P:CAP embeds watermarks independently from the model's main task, which explains why the watermarks cannot be transferred during distillation. 

\subsection{Embedding Robustness against Other Attacks}

\begin{table*}[t]
\centering
\setlength{\tabcolsep}{3.5pt}
\caption{Embedding robustness against other attacks. The host network is trained on MNIST. The classification and watermark accuracies are presented in white and gray columns respectively. Numbers in the 2nd row under P:ING represent the $\lambda$. Numbers in the 2nd column represent the parameters for the corresponding attack.}
\label{tb:mnist_other_attacks}
\begin{tabular}{ll|la|la|la|la|la|la|la|la|la|la}
\hline
\multicolumn{2}{l|}{} & \multicolumn{10}{c|}{Prior Methods} & \multicolumn{10}{c}{P:ING} \\
 \hline
$|D_S|$ & Attacks & \multicolumn{2}{c}{W:LSB} & \multicolumn{2}{c}{W:SGN} & \multicolumn{2}{c}{W:COR} & \multicolumn{2}{c}{W:STA} & \multicolumn{2}{c|}{P:CAP} & \multicolumn{2}{c}{0.5} & \multicolumn{2}{c}{1} & \multicolumn{2}{c}{2} & \multicolumn{2}{c}{4} & \multicolumn{2}{c}{8} \\
\hline
\multirow{3}{*}{1,000} & Pruning(0.4) & 0.97 & 0.24 & 0.98 & 0.53 & 0.97 & 0.42 & 0.98 & 0.77 & 0.98 & 1 & 0.98 & 1 & 0.98 & 0.99 & 0.98 & 0.98 & 0.98 & 0.97 & 0.98 & 0.93 \\
 & Rounding(2) & 0.97 & 0 & 0.98 & 0.75 & 0.97 & 0.41 & 0.99 & 0.8 & 0.98 & 1 & 0.97 & 1 & 0.96 & 1 & 0.95 & 1 & 0.95 & 1 & 0.93 & 1 \\
 & Fine Tuning & 0.99 & 0.27 & 0.99 & 0.59 & 0.99 & 0.48 & 0.98 & 0.8 & 0.98 & 1 & 0.97 & 1 & 0.97 & 1 & 0.97 & 0.99 & 0.97 & 0.98 & 0.97 & 0.95 \\
 \hline
\multirow{3}{*}{5,000} & Pruning(0.4) & 0.98 & 0.23 & 0.98 & 0.52 & 0.98 & 0.42 & 0.98 & 0.77 & 0.98 & 1 & 0.98 & 0.99 & 0.97 & 0.98 & 0.97 & 0.94 & 0.97 & 0.87 & 0.97 & 0.76 \\
 & Rounding(2) & 0.97 & 0 & 0.97 & 0.74 & 0.97 & 0.4 & 0.99 & 0.8 & 0.98 & 1 & 0.96 & 1 & 0.96 & 1 & 0.95 & 1 & 0.94 & 1 & 0.91 & 1 \\
 & Fine Tuning & 0.98 & 0.28 & 0.98 & 0.6 & 0.99 & 0.48 & 0.98 & 0.8 & 0.98 & 1 & 0.97 & 1 & 0.97 & 1 & 0.96 & 0.98 & 0.96 & 0.95 & 0.96 & 0.88 \\
 \hline
\multirow{3}{*}{10,000} & Pruning(0.4) & 0.98 & 0.23 & 0.98 & 0.52 & 0.99 & 0.42 & 0.98 & 0.77 & 0.98 & 1 & 0.98 & 0.99 & 0.98 & 0.96 & 0.97 & 0.89 & 0.97 & 0.79 & 0.97 & 0.64 \\
 & Rounding(2) & 0.97 & 0 & 0.97 & 0.75 & 0.97 & 0.4 & 0.99 & 0.8 & 0.98 & 1 & 0.97 & 1 & 0.96 & 1 & 0.95 & 1 & 0.94 & 1 & 0.9 & 1 \\
 & Fine Tuning & 0.99 & 0.26 & 0.99 & 0.59 & 0.99 & 0.48 & 0.98 & 0.8 & 0.98 & 1 & 0.97 & 1 & 0.97 & 1 & 0.96 & 0.98 & 0.96 & 0.94 & 0.96 & 0.84 \\
 \hline
\end{tabular}
\end{table*}

\begin{table*}[t]
\centering
\setlength{\tabcolsep}{3.3pt}
\caption{Embedding robustness against other attacks. The host network is trained on CIFAR10. The classification and watermark accuracies are presented in white and gray columns respectively. Numbers in the 2nd row under P:ING represent the $\lambda$. Numbers in the 2nd column represent the parameters for the corresponding attack.}
\label{tb:cifar10_other_attacks}
\begin{tabular}{ll|la|la|la|la|la|la|la|la|la|la}
\hline
\multicolumn{2}{l|}{} & \multicolumn{10}{c|}{Prior Methods} & \multicolumn{10}{c}{P:ING} \\
 \hline
$|D_S|$ & Attacks & \multicolumn{2}{c}{W:LSB} & \multicolumn{2}{c}{W:SGN} & \multicolumn{2}{c}{W:COR} & \multicolumn{2}{c}{W:STA} & \multicolumn{2}{c|}{P:CAP} & \multicolumn{2}{c}{0.5} & \multicolumn{2}{c}{1} & \multicolumn{2}{c}{2} & \multicolumn{2}{c}{4} & \multicolumn{2}{c}{8} \\
\hline
\multirow{4}{*}{1,000} & Pruning(0.4) & 0.8 & 0.26 & 0.8 & 0.42 & 0.81 & 0.52 & 0.81 & 0.98 & 0.83 & 1 & 0.82 & 1 & 0.82 & 0.99 & 0.8 & 0.95 & 0.79 & 0.92 & 0.77 & 0.82 \\
 & Rounding(2) & 0.8 & 0.01 & 0.82 & 0.79 & 0.81 & 0.48 & 0.8 & 0.98 & 0.83 & 1 & 0.82 & 1 & 0.81 & 1 & 0.79 & 1 & 0.76 & 1 & 0.71 & 1 \\
 & Fine Tuning & 0.81 & 0.31 & 0.82 & 0.52 & 0.8 & 0.58 & 0.8 & 0.98 & 0.82 & 1 & 0.81 & 1 & 0.79 & 0.98 & 0.79 & 0.98 & 0.77 & 0.9 & 0.75 & 0.83 \\
 & Expansion & 0.75 & 0.65 & 0.76 & 0.64 & 0.75 & 0.63 & 0.78 & 0.88 & 0.79 & 0.97 & 0.79 & 0.97 & 0.78 & 0.97 & 0.76 & 0.97 & 0.73 & 0.97 & 0.67 & 0.96 \\
 \hline
\multirow{4}{*}{5,000} & Pruning(0.4) & 0.79 & 0.25 & 0.81 & 0.41 & 0.79 & 0.51 & 0.81 & 0.98 & 0.82 & 1 & 0.81 & 1 & 0.82 & 1 & 0.81 & 0.99 & 0.8 & 0.93 & 0.78 & 0.76 \\
 & Rounding(2) & 0.8 & 0 & 0.8 & 0.79 & 0.79 & 0.47 & 0.8 & 0.98 & 0.83 & 1 & 0.82 & 1 & 0.81 & 1 & 0.81 & 1 & 0.78 & 1 & 0.73 & 1 \\
 & Fine Tuning & 0.81 & 0.28 & 0.8 & 0.53 & 0.81 & 0.57 & 0.8 & 0.98 & 0.81 & 1 & 0.81 & 1 & 0.8 & 1 & 0.79 & 1 & 0.78 & 0.93 & 0.76 & 0.81 \\
 & Expansion & 0.75 & 0.64 & 0.76 & 0.63 & 0.74 & 0.63 & 0.78 & 0.88 & 0.79 & 0.97 & 0.8 & 0.97 & 0.79 & 0.97 & 0.77 & 0.97 & 0.74 & 0.97 & 0.69 & 0.96 \\
 \hline
\multirow{4}{*}{10,000} & Pruning(0.4) & 0.79 & 0.25 & 0.79 & 0.42 & 0.79 & 0.52 & 0.81 & 0.98 & 0.83 & 0.97 & 0.81 & 0.98 & 0.81 & 0.99 & 0.81 & 0.97 & 0.8 & 0.94 & 0.78 & 0.65 \\
 & Rounding(2) & 0.79 & 0.01 & 0.79 & 0.79 & 0.79 & 0.48 & 0.8 & 0.98 & 0.83 & 1 & 0.82 & 1 & 0.82 & 1 & 0.81 & 1 & 0.79 & 1 & 0.76 & 0.99 \\
 & Fine Tuning & 0.8 & 0.32 & 0.8 & 0.52 & 0.8 & 0.57 & 0.8 & 0.98 & 0.82 & 0.99 & 0.81 & 0.99 & 0.79 & 0.99 & 0.8 & 0.97 & 0.79 & 0.94 & 0.77 & 0.8 \\
 & Expansion & 0.76 & 0.65 & 0.75 & 0.64 & 0.74 & 0.63 & 0.78 & 0.88 & 0.79 & 0.97 & 0.79 & 0.97 & 0.78 & 0.97 & 0.77 & 0.97 & 0.75 & 0.96 & 0.71 & 0.96\\
 \hline
\end{tabular}
\end{table*}

We present the result of robustness evaluation of W:$*$ and P:$*$ methods against other widely used attacks (i.e., pruning, rounding, fine tuning and low-rank expansion) on the MNIST and CIFAR10 classifiers in Table \ref{tb:mnist_other_attacks} and \ref{tb:cifar10_other_attacks} respectively.
The pruning rate is 0.4 and the rounding is by 2 digits.
We do not evaluate the expansion attack on the MNIST classifier because it is a MLP model.

The result shows that, in the prior methods, W:$*$ are generally less robust than P:CAP, which again demonstrates the fragility of W:$*$ methods.
For instance, the watermarks embedded by W:LSB are completely removed by rounding attack and mostly removed by pruning attack.
W:STA embeds the watermarks into the overall statistics of the network parameters as opposed to the individual parameters as in the case of other W:$*$ methods. Therefore, W:STA is slightly more robust than other W:$*$ methods.
The low-rank expansion attack does not touch all the parameters, but rather expands several convolutional layers. 
The watermark information that is embedded in the untouched parameters by W:$*$ can survive the expansion attack. Hence, expansion attacks is less harmful to the embedded watermarks than the other attacks. 
The P:CAP method is quite robust to these attacks compared to W:$*$ method. For example, it achieves 97\%$\sim$100\% watermark accuracy on both classifiers. 

P:ING with $\lambda\leq2$ can achieve comparable watermark accuracy to P:CAP on the MNIST classifier, and on the CIFAR10 classifier, $\lambda\leq4$ leads to comparable watermark accuracy.
This result shows that the robustness of P:ING against other widely used attacks is still comparable to the best result of existing methods.
An interesting finding from the tables is that a larger $\lambda$ of P:ING tends to result in lower watermark accuracy in the pruning and fine tuning attacks.
This is because both attacks require retraining the classifier on a refining set $D'$. 
The retraining process without the govern of the ingrain loss, will drive the classifier to ``forget'' the previously embedded watermarks. 
However, because $D'$ is drawn from the same data distribution $p_x$ as the training set $D$, such effect is negligible unless $\lambda$ becomes large which ingrains the watermarks more deeply in the network.
We present the evaluation result against pruning attack with more rates and rounding attack by more digits in Table \ref{tb:app_prune} and \ref{tb:app_round} respectively in Appendix.

\section{Related Work}
While DL-based systems are reported to attain very high accuracy in various applications~\cite{ML_Go, ML_image}, there remains some limitations that hinder wide adoption of these systems. On the one hand, the computational cost required to operate a neural networks can be prohibitive, often surpassing resource that typically available on mobile devices. On the other hand, various studies have demonstrated fragility of DL models in the presence of adversarial attacks~\cite{defensive_distillation_broken, NN_evasion}. Thus, motivated either by efficiency or security concerns, several model transformation techniques have been studied in the literature~\cite{distilling_DL, defensive_distillation, pruning_quantization}.
Besides model adaptation, various techniques have also been applied in security-related tasks, such as embedding watermark information into a neural network~\cite{embedding_watermarks}, or creating a backdoor in a model~\cite{badnets}.

\subsection{ML Privacy}
Researches have strongly suggested that, similar to other data-driven applications, ML poses a threat to privacy~\cite{attract_info_from_ML, membership_inference_attack, Pharmacogenetics_privacy}. For instance, given access to an ML model, an adversary can infer non-trivial and useful information about its training set~\cite{attract_info_from_ML}. It has also been shown that one can abuse the prediction output by an ML model for an partially unknown input $x$ to infer its unknown features~\cite{Pharmacogenetics_privacy}. Following the same spirit, Shokri et al.~\cite{membership_inference_attack} study membership inference attacks against ML models, wherein an adversary attempts to learn if a record of his choice is part of the private training sets. 

The above mentioned attacks on privacy indicate that benignly trained ML models could leak certain information about its input or training sets. Our study, on the other hand, examines how one can intentionally abuse a model to embed a watermark in a robust way. 

\subsection{Models Abusing}
Model abusing has recently attracted a great attention from the research community. The model abusing problems ask to which extent one can exploit a deep learning model to learn or conduct an additional task beyond the intended task that is associated with the training set~\cite{DL_robust_to_noise, ccs_ML_remembers, embedding_watermarks}. Arpit et al.~\cite{memorizattion_DL} have shown that DL models with sufficient capacity (i.e., having a large enough number of parameters) can ``memorize'' noise contained in training set, without yielding poor generalization to real data samples. Song et al.~\cite{ccs_ML_remembers} explore another abuse which attempts to use the model as a covert channel to convey some secret information (i.e., sensitive information of the training set). 
The authors propose various approaches to achieve this, including poisoning the training set, tampering with the training procedure, or directly modifying parameters of the model after training. 
Motivated by copyright protection of ML models, Uchida et al. \cite{embedding_watermarks} investigated an approach that employs malicious embedding regularizers during training to force the model parameters to observe certain statistical bias, which can then be considered as a ``watermark'' of the model. These abuses have been shown to offer impressive results, strongly indicating that ML models can be abused for additional tasks beyond their intended tasks. 

Nevertheless, it remains unclear if the introduced features are retained after the model undergoes typical ML transformations. Our work, in contrast, experimentally shows that typical ML transformations, especially distillation, are destructive to embedded watermarks which are independent from the model's main task. 

\subsection{Neural Networks in Adversarial Settings}
Deep learning techniques, while achieving utmost accuracy in various application domains~\cite{ML_image_accuracy, ML_Go}, were not originally designed with built-in security. However, recent years have witnessed an ever increasing adoption of DL models for security-sensitive tasks, which causes the accuracy of the models' predictions to have significant implications on the security of the host tasks.
Various works have suggested that ML models are likely vulnerable in adversarial settings~\cite{evadingML_CCS, badnets, defensive_distillation_broken}. In particular, an adversary could force a victim model to deviate from its intended task and behave erratically according to the adversary's wish. The adversary can stage these attacks by corrupting the training phase (e.g., poisoning the training set with adversarial data~\cite{poisoning_SVM, trojanning_attack_NN}, employing adversarial loss function~\cite{badnets}), maliciously modifying the victim model~\cite{defensive_distillation_broken}, or feeding the victim model with adversarially crafted samples~\cite{adversarial_examples_DL, NN_evasion} in the testing phase.

\subsection{Secure $\&$ Privacy-Preserving ML Training}
In the wake of security and privacy threats posed to ML techniques, much research has been devoted to provisioning secure and privacy-preserving training of ML models~\cite{DP_DL, privacypreservingDL, obliviousML}. For instances, Abadi et al. studied a framework to train deep learning models with differential privacy. Shokri et al.~\cite{privacypreservingDL} proposed a protocol for privacy-preserving collaborative deep learning, enables participants to jointly train a model without revealing their private training data. Bonawitz et al.~\cite{secure_aggregation} presented a solution for secure aggregation of high-dimensional data. In addition, systems for oblivious multi-party machine learning have also been built using trusted hardware primitive~\cite{obliviousML}. 

The threat models assumed by these techniques are to protect privacy of users' data contributed to the training set. Our work studies a different threat model wherein the model owner participating in the training process intends to embed watermark in the model.

\subsection{Machine learning vs. Watermarking}
Quiring et al.~\cite{ML_vs_WM} discuss the similarities between machine learning (ML) and watermarking (WM) research. They present two case studies to illustrate such similarities. The first case examines the use of ``1.5-class classifier''  technique in ML (combing two-class and one-class models~\cite{one_and_a_half_class_classifier}) as a defense for oracle attack in WM, whereas the second case study explores the use of stateful detector, which is a common concept in WM, to mitigate model stealing/extraction attacks in ML. 

\section{Conclusion}

We empirically show that all the watermarks embedded by existing methods can be removed by distillation attack.
We design ingrain technique as a countermeasure which mitigates the independence between watermark embedding task and the main task inside the model. Its robustness against other widely used transformations is comparable to existing methods.
Future work includes, but is not limited to, investigating various ways of constructing the ingrainer model, and further enhancing overall embedding robustness against various transformations with minimal loss in the model's main task performance.


\appendix

\begin{table*}[t]
\centering
\setlength{\tabcolsep}{1.8pt}
\caption{Embedding robustness against pruning attack. The classification and watermark accuracies are presented in white and gray columns respectively. Numbers in the 2nd row under P:ING represent the $\lambda$.}
\label{tb:app_prune}
\begin{tabular}{ll|l|la|lalalalala|lalalalala}
\hline
\multicolumn{1}{c}{} & \multicolumn{1}{c}{} & \multicolumn{1}{c}{} & \multicolumn{2}{c|}{} & \multicolumn{10}{c|}{Prior Methods} & \multicolumn{10}{c}{P:ING} \\ \hline
\multicolumn{1}{c}{$|D_S|$} & \multicolumn{1}{c}{$D$} & \multicolumn{1}{c}{Rate} & \multicolumn{2}{c|}{Clean} & \multicolumn{2}{c}{W:LSB} & \multicolumn{2}{c}{W:SGN} & \multicolumn{2}{c}{W:COR} & \multicolumn{2}{c}{W:STA} & \multicolumn{2}{c|}{P:CAP} & \multicolumn{2}{c}{0.5} & \multicolumn{2}{c}{1} & \multicolumn{2}{c}{2} & \multicolumn{2}{c}{4} & \multicolumn{2}{c}{8} \\ \hline
&  & 0.1 & 0.98 & 0.02 & 0.99 & 0.27 & 0.99 & 0.59 & 0.99 & 0.48 & 0.98 & 0.8 & \textbf{0.99} & \textbf{1} & \textbf{0.98} & \textbf{1} & \textbf{0.98} & \textbf{0.99} & \textbf{0.98} & \textbf{0.99} & \textbf{0.98} & \textbf{0.96} & \textbf{0.98} & \textbf{0.93} \\
  &  & 0.2 & 0.98 & 0.01 & 0.98 & 0.29 & 0.99 & 0.6 & 0.98 & 0.49 & 0.98 & 0.8 & \textbf{0.98} & \textbf{1} & \textbf{0.98} & \textbf{1} & \textbf{0.98} & \textbf{1} & \textbf{0.98} & \textbf{0.98} & \textbf{0.98} & \textbf{0.97} & \textbf{0.98} & \textbf{0.93} \\
  &  & 0.3 & 0.99 & 0.02 & 0.97 & 0.26 & 0.98 & 0.56 & 0.97 & 0.45 & 0.98 & 0.8 & \textbf{0.98} & \textbf{1} & \textbf{0.98} & \textbf{1} & \textbf{0.98} & \textbf{1} & \textbf{0.98} & \textbf{0.98} & \textbf{0.98} & \textbf{0.97} & \textbf{0.98} & \textbf{0.93} \\
  &  & 0.4 & 0.99 & 0.01 & 0.97 & 0.24 & 0.98 & 0.53 & 0.97 & 0.42 & 0.98 & 0.77 & \textbf{0.98} & \textbf{1} & \textbf{0.98} & \textbf{1} & \textbf{0.98} & \textbf{0.99} & \textbf{0.98} & \textbf{0.98} & \textbf{0.98} & \textbf{0.97} & \textbf{0.98} & \textbf{0.93} \\
 1,000  & MNIST & 0.5 & 0.99 & 0.02 & 0.98 & 0.12 & 0.98 & 0.45 & 0.98 & 0.37 & 0.98 & 0.74 & \textbf{0.98} & \textbf{1} & \textbf{0.98} & \textbf{0.99} & \textbf{0.98} & \textbf{0.98} & \textbf{0.98} & \textbf{0.97} & \textbf{0.98} & \textbf{0.95} & \textbf{0.98} & \textbf{0.91} \\
  &  & 0.6 & 0.99 & 0.02 & 0.98 & 0.05 & 0.98 & 0.42 & 0.98 & 0.37 & 0.98 & 0.69 & \textbf{0.98} & \textbf{1} & \textbf{0.98} & \textbf{0.98} & \textbf{0.98} & \textbf{0.96} & \textbf{0.98} & \textbf{0.93} & \textbf{0.98} & \textbf{0.91} & 0.98 & 0.88 \\
     \hhline{~|-|-|-|-|-|-|-|-|-|-|-|-|-|-|-|-|-|-|-|-|-|-|-|-|}
  &  & 0.1 & 0.83 & 0.01 & 0.81 & 0.31 & 0.82 & 0.52 & 0.8 & 0.58 & \textbf{0.82} & \textbf{0.98} & \textbf{0.82} & \textbf{1} & \textbf{0.82} & \textbf{1} & \textbf{0.82} & \textbf{0.98} & \textbf{0.8} & \textbf{0.95} & 0.79 & 0.88 & 0.77 & 0.79 \\
  &  & 0.2 & 0.82 & 0 & 0.81 & 0.35 & 0.81 & 0.56 & 0.82 & 0.63 & \textbf{0.82} & \textbf{0.98} & \textbf{0.82} & \textbf{1} & \textbf{0.82} & \textbf{1} & \textbf{0.82} & \textbf{0.99} & \textbf{0.81} & \textbf{0.93} & 0.79 & 0.9 & 0.77 & 0.79 \\
  &  & 0.3 & 0.83 & 0.01 & 0.8 & 0.32 & 0.81 & 0.52 & 0.82 & 0.58 & \textbf{0.81} & \textbf{0.98} & \textbf{0.82} & \textbf{1} & \textbf{0.82} & \textbf{1} & \textbf{0.81} & \textbf{0.98} & \textbf{0.81} & \textbf{0.95} & 0.79 & 0.91 & 0.76 & 0.76 \\
  &  & 0.4 & 0.83 & 0 & 0.8 & 0.26 & 0.8 & 0.42 & 0.81 & 0.52 & \textbf{0.81} & \textbf{0.98} & \textbf{0.83} & \textbf{1} & \textbf{0.82} & \textbf{1} & \textbf{0.82} & \textbf{0.99} & \textbf{0.8} & \textbf{0.95} & 0.79 & 0.92 & 0.77 & 0.82 \\
 & CIFAR10 & 0.5 & 0.83 & 0 & 0.81 & 0.06 & 0.81 & 0.3 & 0.81 & 0.41 & \textbf{0.82} & \textbf{0.98} & \textbf{0.82} & \textbf{1} & \textbf{0.82} & \textbf{1} & \textbf{0.81} & \textbf{0.99} & \textbf{0.8} & \textbf{0.95} & 0.79 & 0.91 & 0.76 & 0.77 \\
  &  & 0.6 & 0.83 & 0.01 & 0.81 & 0 & 0.8 & 0.36 & 0.8 & 0.38 & \textbf{0.82} & \textbf{0.98} & \textbf{0.82} & \textbf{1} & \textbf{0.82} & \textbf{1} & \textbf{0.82} & \textbf{0.99} & \textbf{0.8} & \textbf{0.96} & 0.79 & 0.89 & 0.77 & 0.78 \\
\hline
 &  & 0.1 & 0.98 & 0 & 0.98 & 0.28 & 0.98 & 0.6 & 0.99 & 0.48 & 0.98 & 0.8 & \textbf{0.98} & \textbf{1} & \textbf{0.98} & \textbf{1} & \textbf{0.97} & \textbf{0.99} & \textbf{0.97} & \textbf{0.96} & \textbf{0.97} & \textbf{0.91} & 0.97 & 0.81 \\
  &  & 0.2 & 0.98 & 0.01 & 0.98 & 0.3 & 0.98 & 0.6 & 0.98 & 0.5 & 0.98 & 0.8 & \textbf{0.98} & \textbf{1} & \textbf{0.98} & \textbf{1} & \textbf{0.97} & \textbf{0.99} & \textbf{0.97} & \textbf{0.96} & \textbf{0.97} & \textbf{0.91} & 0.97 & 0.81 \\
  &  & 0.3 & 0.99 & 0 & 0.98 & 0.27 & 0.97 & 0.57 & 0.94 & 0.47 & 0.98 & 0.8 & \textbf{0.98} & \textbf{1} & \textbf{0.98} & \textbf{1} & \textbf{0.97} & \textbf{0.99} & \textbf{0.97} & \textbf{0.96} & \textbf{0.97} & \textbf{0.91} & 0.97 & 0.79 \\
  &  & 0.4 & 0.99 & 0.01 & 0.98 & 0.23 & 0.98 & 0.52 & 0.98 & 0.42 & 0.98 & 0.77 & \textbf{0.98} & \textbf{1} & \textbf{0.98} & \textbf{0.99} & \textbf{0.97} & \textbf{0.98} & \textbf{0.97} & \textbf{0.94} & 0.97 & 0.87 & 0.97 & 0.76 \\
 5,000  & MNIST & 0.5 & 0.99 & 0 & 0.99 & 0.13 & 0.97 & 0.44 & 0.98 & 0.36 & 0.98 & 0.74 & \textbf{0.98} & \textbf{1} & \textbf{0.98} & \textbf{0.98} & \textbf{0.97} & \textbf{0.95} & 0.97 & 0.89 & 0.97 & 0.81 & 0.97 & 0.69 \\
  &  & 0.6 & 0.99 & 0 & 0.98 & 0.05 & 0.97 & 0.41 & 0.98 & 0.37 & 0.98 & 0.69 & \textbf{0.98} & \textbf{1} & \textbf{0.98} & \textbf{0.93} & 0.98 & 0.86 & 0.97 & 0.79 & 0.97 & 0.69 & 0.97 & 0.57 \\
     \hhline{~|-|-|-|-|-|-|-|-|-|-|-|-|-|-|-|-|-|-|-|-|-|-|-|-|}
  &  & 0.1 & 0.83 & 0.01 & 0.81 & 0.28 & 0.8 & 0.53 & 0.81 & 0.57 & \textbf{0.82} & \textbf{0.98} & \textbf{0.82} & \textbf{1} & \textbf{0.82} & \textbf{1} & \textbf{0.82} & \textbf{0.99} & \textbf{0.81} & \textbf{0.93} & \textbf{0.8} & \textbf{0.91} & 0.78 & 0.7 \\
  &  & 0.2 & 0.82 & 0 & 0.8 & 0.35 & 0.81 & 0.55 & 0.8 & 0.62 & \textbf{0.82} & \textbf{0.98} & \textbf{0.82} & \textbf{1} & \textbf{0.82} & \textbf{1} & \textbf{0.82} & \textbf{1} & \textbf{0.82} & \textbf{0.97} & \textbf{0.81} & \textbf{0.93} & 0.78 & 0.75 \\
  &  & 0.3 & 0.83 & 0.01 & 0.8 & 0.34 & 0.8 & 0.54 & 0.8 & 0.58 & \textbf{0.81} & \textbf{0.98} & \textbf{0.82} & \textbf{1} & \textbf{0.82} & \textbf{1} & \textbf{0.82} & \textbf{1} & \textbf{0.81} & \textbf{0.99} & \textbf{0.8} & \textbf{0.94} & 0.78 & 0.76 \\
  &  & 0.4 & 0.83 & 0 & 0.79 & 0.25 & 0.81 & 0.41 & 0.79 & 0.51 & \textbf{0.81} & \textbf{0.98} & \textbf{0.82} & \textbf{1} & \textbf{0.81} & \textbf{1} & \textbf{0.82} & \textbf{1} & \textbf{0.81} & \textbf{0.99} & \textbf{0.8} & \textbf{0.93} & 0.78 & 0.76 \\
 & CIFAR10 & 0.5 & 0.83 & 0 & 0.79 & 0.06 & 0.81 & 0.3 & 0.81 & 0.41 & \textbf{0.82} & \textbf{0.98} & \textbf{0.82} & \textbf{1} & \textbf{0.81} & \textbf{1} & \textbf{0.82} & \textbf{1} & \textbf{0.82} & \textbf{0.96} & \textbf{0.8} & \textbf{0.93} & 0.77 & 0.75 \\
  &  & 0.6 & 0.83 & 0 & 0.79 & 0 & 0.81 & 0.35 & 0.8 & 0.38 & \textbf{0.82} & \textbf{0.98} & \textbf{0.82} & \textbf{1} & \textbf{0.82} & \textbf{1} & \textbf{0.82} & \textbf{0.99} & \textbf{0.81} & \textbf{0.93} & 0.8 & 0.88 & 0.78 & 0.68 \\
\hline
  &  & 0.1 & 0.98 & 0 & 0.99 & 0.26 & 0.99 & 0.59 & 0.99 & 0.48 & 0.98 & 0.8 & \textbf{0.98} & \textbf{1} & \textbf{0.98} & \textbf{1} & \textbf{0.98} & \textbf{0.99} & \textbf{0.97} & \textbf{0.94} & 0.97 & 0.84 & 0.97 & 0.7 \\
  &  & 0.2 & 0.98 & 0 & 0.98 & 0.3 & 0.98 & 0.6 & 0.98 & 0.49 & 0.98 & 0.8 & \textbf{0.98} & \textbf{1} & \textbf{0.98} & \textbf{1} & \textbf{0.98} & \textbf{0.99} & \textbf{0.97} & \textbf{0.94} & 0.97 & 0.85 & 0.97 & 0.7 \\
  &  & 0.3 & 0.99 & 0 & 0.95 & 0.26 & 0.96 & 0.56 & 0.96 & 0.45 & 0.98 & 0.8 & \textbf{0.98} & \textbf{1} & \textbf{0.98} & \textbf{1} & \textbf{0.98} & \textbf{0.98} & \textbf{0.97} & \textbf{0.93} & 0.97 & 0.83 & 0.97 & 0.68 \\
  &  & 0.4 & 0.99 & 0 & 0.98 & 0.23 & 0.98 & 0.52 & 0.99 & 0.42 & 0.98 & 0.77 & \textbf{0.98} & \textbf{1} & \textbf{0.98} & \textbf{0.99} & \textbf{0.98} & \textbf{0.96} & \textbf{0.97} & \textbf{0.89} & 0.97 & 0.79 & 0.97 & 0.64 \\
 10,000  & MNIST & 0.5 & 0.99 & 0 & 0.98 & 0.12 & 0.98 & 0.44 & 0.98 & 0.37 & 0.98 & 0.74 & \textbf{0.98} & \textbf{1} & \textbf{0.98} & \textbf{0.96} & \textbf{0.98} & \textbf{0.9} & 0.97 & 0.81 & 0.97 & 0.7 & 0.97 & 0.55 \\
  &  & 0.6 & 0.99 & 0 & 0.98 & 0.05 & 0.98 & 0.41 & 0.98 & 0.37 & 0.98 & 0.69 & \textbf{0.98} & \textbf{0.99} & \textbf{0.98} & \textbf{0.86} & 0.98 & 0.77 & 0.97 & 0.66 & 0.97 & 0.55 & 0.97 & 0.44 \\
     \hhline{~|-|-|-|-|-|-|-|-|-|-|-|-|-|-|-|-|-|-|-|-|-|-|-|-|}
 
  &  & 0.1 & 0.83 & 0 & 0.8 & 0.32 & 0.8 & 0.52 & 0.8 & 0.57 & \textbf{0.82} & \textbf{0.98} & \textbf{0.82} & \textbf{0.95} & \textbf{0.82} & \textbf{0.97} & \textbf{0.81} & \textbf{0.97} & \textbf{0.81} & \textbf{0.97} & \textbf{0.81} & \textbf{0.92} & 0.79 & 0.63 \\
  &  & 0.2 & 0.82 & 0 & 0.8 & 0.35 & 0.8 & 0.56 & 0.8 & 0.63 & \textbf{0.82} & \textbf{0.98} & \textbf{0.82} & \textbf{0.94} & \textbf{0.81} & \textbf{0.95} & \textbf{0.81} & \textbf{0.98} & \textbf{0.81} & \textbf{0.97} & \textbf{0.81} & \textbf{0.93} & 0.79 & 0.68 \\
  &  & 0.3 & 0.83 & 0 & 0.79 & 0.32 & 0.8 & 0.54 & 0.79 & 0.58 & \textbf{0.81} & \textbf{0.98} & \textbf{0.82} & \textbf{0.97} & \textbf{0.81} & \textbf{0.99} & \textbf{0.81} & \textbf{0.98} & \textbf{0.81} & \textbf{0.97} & \textbf{0.81} & \textbf{0.93} & 0.78 & 0.66 \\
  &  & 0.4 & 0.83 & 0 & 0.79 & 0.25 & 0.79 & 0.42 & 0.79 & 0.52 & \textbf{0.81} & \textbf{0.98} & \textbf{0.83} & \textbf{0.97} & \textbf{0.81} & \textbf{0.98} & \textbf{0.81} & \textbf{0.99} & \textbf{0.81} & \textbf{0.97} & \textbf{0.8} & \textbf{0.94} & 0.78 & 0.65 \\
 & CIFAR10 & 0.5 & 0.83 & 0 & 0.8 & 0.06 & 0.79 & 0.3 & 0.78 & 0.41 & \textbf{0.82} & \textbf{0.98} & \textbf{0.83} & \textbf{0.96} & \textbf{0.82} & \textbf{0.96} & \textbf{0.82} & \textbf{0.98} & \textbf{0.81} & \textbf{0.98} & \textbf{0.81} & \textbf{0.94} & 0.79 & 0.63 \\
  &  & 0.6 & 0.83 & 0 & 0.79 & 0 & 0.78 & 0.35 & 0.8 & 0.37 & \textbf{0.82} & \textbf{0.98} & \textbf{0.83} & \textbf{0.95} & \textbf{0.82} & \textbf{0.97} & \textbf{0.82} & \textbf{0.96} & \textbf{0.81} & \textbf{0.94} & \textbf{0.8} & \textbf{0.9} & 0.79 & 0.6 \\
  \hline
 \end{tabular}
\end{table*}

\begin{table*}[]
\setlength{\tabcolsep}{1.5pt}
\centering
\caption{Embedding robustness against rounding attack. The classification and watermark accuracies are presented in white and gray columns respectively. Numbers in the 2nd row under P:ING represent the $\lambda$.}
\label{tb:app_round}
\begin{tabular}{ll|l|la|lalalalala|lalalalala}
\hline
\multicolumn{1}{c}{} & \multicolumn{1}{c}{} & \multicolumn{1}{c}{} & \multicolumn{2}{c|}{} & \multicolumn{10}{c|}{Prior Methods} & \multicolumn{10}{c}{P:ING} \\ \hline
\multicolumn{1}{c}{$|D_S|$} & \multicolumn{1}{c}{$D$} & \multicolumn{1}{c}{\#Digits} & \multicolumn{2}{c|}{Clean} & \multicolumn{2}{c}{W:LSB} & \multicolumn{2}{c}{W:SGN} & \multicolumn{2}{c}{W:COR} & \multicolumn{2}{c}{W:STA} & \multicolumn{2}{c|}{P:CAP} & \multicolumn{2}{c}{0.5} & \multicolumn{2}{c}{1} & \multicolumn{2}{c}{2} & \multicolumn{2}{c}{4} & \multicolumn{2}{c}{8} \\ \hline
&  & 1 & 0.97 & 0.01 & 0.96 & 0.02 & 0.97 & 0.75 & 0.97 & 0.24 & 0.97 & 0.41 & \textbf{0.97} & \textbf{0.88} & \textbf{0.78} & \textbf{0.92} & \textbf{0.62} & \textbf{0.92} & \textbf{0.45} & \textbf{0.92} & \textbf{0.45} & \textbf{0.89} & \textbf{0.44} & \textbf{0.91} \\
  &  & 2 & 0.98 & 0.02 & 0.97 & 0 & 0.98 & 0.75 & 0.97 & 0.41 & 0.99 & 0.8 & \textbf{0.98} & \textbf{1} & \textbf{0.97} & \textbf{1} & \textbf{0.96} & \textbf{1} & \textbf{0.95} & \textbf{1} & \textbf{0.95} & \textbf{1} & \textbf{0.93} & \textbf{1} \\
 1,000  & MNIST & 3 & 0.98 & 0.02 & 0.98 & 0.07 & 0.98 & 0.75 & 0.98 & 0.52 & 0.98 & 0.8 & \textbf{0.98} & \textbf{1} & \textbf{0.97} & \textbf{1} & \textbf{0.96} & \textbf{1} & \textbf{0.96} & \textbf{1} & \textbf{0.95} & \textbf{1} & \textbf{0.93} & \textbf{1} \\
  &  & 4 & 0.98 & 0.02 & 0.98 & 0.11 & 0.98 & 0.75 & 0.98 & 0.56 & 0.98 & 0.8 & \textbf{0.98} & \textbf{1} & \textbf{0.97} & \textbf{1} & \textbf{0.96} & \textbf{1} & \textbf{0.95} & \textbf{1} & \textbf{0.95} & \textbf{1} & \textbf{0.93} & \textbf{1} \\
  &  & 5 & 0.98 & 0.02 & 0.98 & 0.17 & 0.98 & 0.75 & 0.98 & 0.59 & 0.98 & 0.8 & \textbf{0.98} & \textbf{1} & \textbf{0.97} & \textbf{1} & \textbf{0.96} & \textbf{1} & \textbf{0.95} & \textbf{1} & \textbf{0.95} & \textbf{1} & \textbf{0.93} & \textbf{1} \\
  &  & 6 & 0.98 & 0.02 & 0.98 & 0.25 & 0.98 & 0.75 & 0.98 & 0.72 & 0.98 & 0.8 & \textbf{0.98} & \textbf{1} & \textbf{0.97} & \textbf{1} & \textbf{0.96} & \textbf{1} & \textbf{0.95} & \textbf{1} & \textbf{0.95} & \textbf{1} & \textbf{0.93} & \textbf{1} \\
    \hhline{~|-|-|-|-|-|-|-|-|-|-|-|-|-|-|-|-|-|-|-|-|-|-|-|-|}
 &  & 1 & 0.79 & 0.02 & 0.75 & 0 & 0.74 & 0.79 & 0.74 & 0.24 & \textbf{0.76} & \textbf{0.96} & \textbf{0.8} & \textbf{0.93} & 0.75 & 0.88 & \textbf{0.74} & \textbf{0.95} & \textbf{0.75} & \textbf{0.95} & \textbf{0.72} & \textbf{0.97} & \textbf{0.69} & \textbf{0.97} \\
  &  & 2 & 0.83 & 0 & 0.8 & 0.01 & 0.82 & 0.79 & 0.81 & 0.48 & \textbf{0.8} & \textbf{0.98} & \textbf{0.83} & \textbf{1} & \textbf{0.82} & \textbf{1} & \textbf{0.81} & \textbf{1} & \textbf{0.79} & \textbf{1} & \textbf{0.76} & \textbf{1} & \textbf{0.71} & \textbf{1} \\
 & CIFAR10 & 3 & 0.83 & 0 & 0.8 & 0.09 & 0.8 & 0.79 & 0.82 & 0.54 & \textbf{0.8} & \textbf{0.98} & \textbf{0.83} & \textbf{1} & \textbf{0.82} & \textbf{1} & \textbf{0.81} & \textbf{1} & \textbf{0.79} & \textbf{1} & \textbf{0.76} & \textbf{1} & \textbf{0.71} & \textbf{1} \\
  &  & 4 & 0.84 & 0 & 0.81 & 0.13 & 0.81 & 0.79 & 0.8 & 0.59 & \textbf{0.8} & \textbf{0.98} & \textbf{0.83} & \textbf{1} & \textbf{0.82} & \textbf{1} & \textbf{0.81} & \textbf{1} & \textbf{0.79} & \textbf{1} & \textbf{0.76} & \textbf{1} & \textbf{0.71} & \textbf{1} \\
  &  & 5 & 0.84 & 0 & 0.81 & 0.19 & 0.81 & 0.79 & 0.81 & 0.63 & \textbf{0.8} & \textbf{0.98} & \textbf{0.83} & \textbf{1} & \textbf{0.82} & \textbf{1} & \textbf{0.81} & \textbf{1} & \textbf{0.79} & \textbf{1} & \textbf{0.76} & \textbf{1} & \textbf{0.71} & \textbf{1} \\
  &  & 6 & 0.84 & 0 & 0.81 & 0.31 & 0.8 & 0.79 & 0.8 & 0.77 & \textbf{0.8} & \textbf{0.98} & \textbf{0.83} & \textbf{1} & \textbf{0.82} & \textbf{1} & \textbf{0.81} & \textbf{1} & \textbf{0.79} & \textbf{1} & \textbf{0.76} & \textbf{1} & \textbf{0.71} & \textbf{1} \\
\hline
&  & 1 & 0.97 & 0.01 & 0.95 & 0.02 & \textbf{0.96} & \textbf{0.74} & 0.97 & 0.23 & 0.97 & 0.41 & \textbf{0.95} & \textbf{0.68} & \textbf{0.77} & \textbf{0.66} & \textbf{0.67} & \textbf{0.6} & 0.54 & 0.58 & 0.53 & 0.55 & 0.53 & 0.52 \\
  &  & 2 & 0.98 & 0 & 0.97 & 0 & 0.97 & 0.74 & 0.97 & 0.4 & 0.99 & 0.8 & \textbf{0.98} & \textbf{1} & \textbf{0.96} & \textbf{1} & \textbf{0.96} & \textbf{1} & \textbf{0.95} & \textbf{1} & \textbf{0.94} & \textbf{1} & \textbf{0.91} & \textbf{1} \\
 5,000  & MNIST & 3 & 0.98 & 0 & 0.98 & 0.07 & 0.97 & 0.74 & 0.97 & 0.52 & 0.98 & 0.8 & \textbf{0.98} & \textbf{1} & \textbf{0.96} & \textbf{1} & \textbf{0.96} & \textbf{1} & \textbf{0.95} & \textbf{1} & \textbf{0.94} & \textbf{1} & \textbf{0.92} & \textbf{1} \\
  &  & 4 & 0.98 & 0 & 0.98 & 0.11 & 0.98 & 0.74 & 0.98 & 0.56 & 0.98 & 0.8 & \textbf{0.98} & \textbf{1} & \textbf{0.96} & \textbf{1} & \textbf{0.96} & \textbf{1} & \textbf{0.95} & \textbf{1} & \textbf{0.94} & \textbf{1} & \textbf{0.92} & \textbf{1} \\
  &  & 5 & 0.98 & 0 & 0.98 & 0.17 & 0.98 & 0.74 & 0.97 & 0.58 & 0.98 & 0.8 & \textbf{0.98} & \textbf{1} & \textbf{0.96} & \textbf{1} & \textbf{0.96} & \textbf{1} & \textbf{0.95} & \textbf{1} & \textbf{0.94} & \textbf{1} & \textbf{0.92} & \textbf{1} \\
  &  & 6 & 0.98 & 0.01 & 0.98 & 0.25 & 0.98 & 0.74 & 0.98 & 0.72 & 0.98 & 0.8 & \textbf{0.98} & \textbf{1} & \textbf{0.96} & \textbf{1} & \textbf{0.96} & \textbf{1} & \textbf{0.95} & \textbf{1} & \textbf{0.94} & \textbf{1} & \textbf{0.92} & \textbf{1} \\
    \hhline{~|-|-|-|-|-|-|-|-|-|-|-|-|-|-|-|-|-|-|-|-|-|-|-|-|}
  &  & 1 & 0.79 & 0.02 & 0.76 & 0 & 0.76 & 0.79 & 0.76 & 0.24 & \textbf{0.76} & \textbf{0.96} & \textbf{0.79} & \textbf{0.9} & \textbf{0.76} & \textbf{0.92} & \textbf{0.74} & \textbf{0.89} & \textbf{0.73} & \textbf{0.93} & \textbf{0.72} & \textbf{0.95} & \textbf{0.7} & \textbf{0.91} \\
  &  & 2 & 0.83 & 0 & 0.8 & 0 & 0.8 & 0.79 & 0.79 & 0.47 & \textbf{0.8} & \textbf{0.98} & \textbf{0.83} & \textbf{1} & \textbf{0.82} & \textbf{1} & \textbf{0.81} & \textbf{1} & \textbf{0.81} & \textbf{1} & \textbf{0.78} & \textbf{1} & \textbf{0.73} & \textbf{1} \\
 & CIFAR10 & 3 & 0.83 & 0 & 0.8 & 0.09 & 0.81 & 0.79 & 0.79 & 0.54 & \textbf{0.8} & \textbf{0.98} & \textbf{0.83} & \textbf{1} & \textbf{0.82} & \textbf{1} & \textbf{0.82} & \textbf{1} & \textbf{0.81} & \textbf{1} & \textbf{0.78} & \textbf{1} & \textbf{0.73} & \textbf{1} \\
  &  & 4 & 0.84 & 0 & 0.8 & 0.12 & 0.81 & 0.79 & 0.79 & 0.58 & \textbf{0.8} & \textbf{0.98} & \textbf{0.83} & \textbf{1} & \textbf{0.82} & \textbf{1} & \textbf{0.82} & \textbf{1} & \textbf{0.81} & \textbf{1} & \textbf{0.78} & \textbf{1} & \textbf{0.73} & \textbf{1} \\
  &  & 5 & 0.84 & 0 & 0.8 & 0.19 & 0.81 & 0.79 & 0.8 & 0.63 & \textbf{0.8} & \textbf{0.98} & \textbf{0.83} & \textbf{1} & \textbf{0.82} & \textbf{1} & \textbf{0.82} & \textbf{1} & \textbf{0.81} & \textbf{1} & \textbf{0.78} & \textbf{1} & \textbf{0.73} & \textbf{1} \\
  &  & 6 & 0.84 & 0 & 0.8 & 0.3 & 0.81 & 0.79 & 0.8 & 0.76 & \textbf{0.8} & \textbf{0.98} & \textbf{0.83} & \textbf{1} & \textbf{0.82} & \textbf{1} & \textbf{0.82} & \textbf{1} & \textbf{0.81} & \textbf{1} & \textbf{0.78} & \textbf{1} & \textbf{0.73} & \textbf{1} \\
\hline
&  & 1 & 0.97 & 0 & 0.92 & 0.02 & \textbf{0.91} & \textbf{0.75} & 0.9 & 0.24 & 0.97 & 0.41 & \textbf{0.92} & \textbf{0.51} & \textbf{0.79} & \textbf{0.45} & \textbf{0.62} & \textbf{0.44} & \textbf{0.55} & \textbf{0.43} & 0.57 & 0.4 & 0.56 & 0.38 \\
  &  & 2 & 0.98 & 0 & 0.97 & 0 & 0.97 & 0.75 & 0.97 & 0.4 & 0.99 & 0.8 & \textbf{0.98} & \textbf{1} & \textbf{0.97} & \textbf{1} & \textbf{0.96} & \textbf{1} & \textbf{0.95} & \textbf{1} & \textbf{0.94} & \textbf{1} & \textbf{0.9} & \textbf{1} \\
 10,000  & MNIST & 3 & 0.98 & 0 & 0.97 & 0.07 & 0.97 & 0.75 & 0.97 & 0.51 & 0.98 & 0.8 & \textbf{0.98} & \textbf{1} & \textbf{0.97} & \textbf{1} & \textbf{0.96} & \textbf{1} & \textbf{0.95} & \textbf{1} & \textbf{0.94} & \textbf{1} & \textbf{0.91} & \textbf{1} \\
  &  & 4 & 0.98 & 0 & 0.97 & 0.11 & 0.97 & 0.75 & 0.98 & 0.56 & 0.98 & 0.8 & \textbf{0.98} & \textbf{1} & \textbf{0.97} & \textbf{1} & \textbf{0.96} & \textbf{1} & \textbf{0.95} & \textbf{1} & \textbf{0.94} & \textbf{1} & \textbf{0.91} & \textbf{1} \\
  &  & 5 & 0.98 & 0 & 0.97 & 0.17 & 0.97 & 0.75 & 0.98 & 0.59 & 0.98 & 0.8 & \textbf{0.98} & \textbf{1} & \textbf{0.97} & \textbf{1} & \textbf{0.96} & \textbf{1} & \textbf{0.95} & \textbf{1} & \textbf{0.94} & \textbf{1} & \textbf{0.91} & \textbf{1} \\
  &  & 6 & 0.98 & 0 & 0.98 & 0.25 & 0.98 & 0.75 & 0.98 & 0.72 & 0.98 & 0.8 & \textbf{0.98} & \textbf{1} & \textbf{0.97} & \textbf{1} & \textbf{0.96} & \textbf{1} & \textbf{0.95} & \textbf{1} & \textbf{0.94} & \textbf{1} & \textbf{0.91} & \textbf{1} \\
    \hhline{~|-|-|-|-|-|-|-|-|-|-|-|-|-|-|-|-|-|-|-|-|-|-|-|-|}
  &  & 1 & 0.79 & 0 & 0.75 & 0 & 0.76 & 0.79 & 0.76 & 0.23 & \textbf{0.76} & \textbf{0.96} & \textbf{0.8} & \textbf{0.72} & \textbf{0.73} & \textbf{0.74} & \textbf{0.75} & \textbf{0.83} & \textbf{0.71} & \textbf{0.85} & \textbf{0.7} & \textbf{0.87} & 0.69 & 0.68 \\
  &  & 2 & 0.83 & 0 & 0.79 & 0.01 & 0.79 & 0.79 & 0.79 & 0.48 & \textbf{0.8} & \textbf{0.98} & \textbf{0.83} & \textbf{1} & \textbf{0.82} & \textbf{1} & \textbf{0.82} & \textbf{1} & \textbf{0.81} & \textbf{1} & \textbf{0.79} & \textbf{1} & \textbf{0.76} & \textbf{0.99} \\
 & CIFAR10 & 3 & 0.83 & 0 & 0.79 & 0.09 & 0.79 & 0.79 & 0.79 & 0.54 & \textbf{0.8} & \textbf{0.98} & \textbf{0.83} & \textbf{1} & \textbf{0.82} & \textbf{1} & \textbf{0.82} & \textbf{1} & \textbf{0.81} & \textbf{1} & \textbf{0.79} & \textbf{1} & \textbf{0.76} & \textbf{0.99} \\
  &  & 4 & 0.84 & 0 & 0.79 & 0.13 & 0.8 & 0.79 & 0.79 & 0.58 & \textbf{0.8} & \textbf{0.98} & \textbf{0.83} & \textbf{1} & \textbf{0.82} & \textbf{1} & \textbf{0.82} & \textbf{1} & \textbf{0.81} & \textbf{1} & \textbf{0.79} & \textbf{1} & \textbf{0.76} & \textbf{0.99} \\
  &  & 5 & 0.84 & 0 & 0.8 & 0.19 & 0.8 & 0.79 & 0.78 & 0.63 & \textbf{0.8} & \textbf{0.98} & \textbf{0.83} & \textbf{1} & \textbf{0.82} & \textbf{1} & \textbf{0.82} & \textbf{1} & \textbf{0.81} & \textbf{1} & \textbf{0.79} & \textbf{1} & \textbf{0.76} & \textbf{0.99} \\
  &  & 6 & 0.84 & 0 & 0.8 & 0.3 & 0.78 & 0.79 & 0.78 & 0.77 & \textbf{0.8} & \textbf{0.98} & \textbf{0.83} & \textbf{1} & \textbf{0.82} & \textbf{1} & \textbf{0.82} & \textbf{1} & \textbf{0.81} & \textbf{1} & \textbf{0.79} & \textbf{1} & \textbf{0.76} & \textbf{0.99} \\
\hline
 \end{tabular}
\end{table*}

\end{document}